\documentclass[useAMS,usenatbib]{mn2e}

\usepackage{amsmath}
\usepackage{lipsum} % for dummy text only
\usepackage{graphicx}
\usepackage{amssymb}
\usepackage{color}
\usepackage{hyperref}
\hypersetup{pdftex,  % needed for pdflatex
  breaklinks=true,  % so long urls are correctly broken across lines
  colorlinks=true,
  urlcolor=blue,
  linkcolor=darkorange,
  citecolor=darkgreen,
  }

\newcommand\nodata{ ~$\cdots$~ }%

\definecolor{orange}{cmyk}{0,0.4,0.8,0.2}
\definecolor{darkorange}{rgb}{.71,0.21,0.01}
\definecolor{darkgreen}{rgb}{.12,.54,.11}

\newcommand{\citepeg}[1]{\citep[{e.g.,}][]{#1}}

\newcommand\cutinhead[2]{}

\renewcommand\cutinhead[2]{%                                                      
 \noalign{\vskip 1.5ex}%                                                        
 \hline
\\
 \noalign{\vskip -1.5ex}%                                                       
 \multicolumn{#2}{c}{#1}%                                                 
\\
 \noalign{\vskip .8ex}%                                                         
 \hline
\\
 \noalign{\vskip -2ex}%                                                         
}%

\title[Multi-band RR Lyrae period--magnitude relations]{Towards precision distances and 3D dust maps using  broadband Period--Magnitude relations of RR Lyrae stars}
\author[Klein \& Bloom]{C. R. Klein$^{1}$\thanks{E-mail:
cklein@berkeley.edu} and J. S. Bloom$^{1,2}$\\
$^{1}$Astronomy Department, University of California, Berkeley, CA 94720, USA \\
$^{2}$Physics Division, 1 Cyclotron Road, Lawrence Berkeley National Laboratory, CA 94720, USA}
\begin{document}

\date{Submitted 2014 April 18.}

\pagerange{\pageref{firstpage}--\pageref{lastpage}} \pubyear{2014}

\maketitle

\label{firstpage}

\begin{abstract}
We determine the period-magnitude relations of RR Lyrae stars in 13 photometric bandpasses from 0.4 to 12 $\mu$m using timeseries observations of 134 stars with prior parallax measurements from {\it Hipparcos} and the {\it Hubble Space Telescope} (HST). The Bayesian formalism, extended from our previous work to include the effects of line-of-sight dust extinction, allows for the simultaneous inference of the posterior distribution of the mean absolute magnitude, slope of the period-magnitude power-law, and intrinsic scatter about a perfect power-law for each bandpass. In addition, the distance modulus and line-of-sight dust extinction to each RR Lyrae star in the calibration sample is determined, yielding a sample median fractional distance error of 0.66 per cent. The intrinsic scatter in all bands appears to be larger than the photometric errors, except in {\it WISE} $W1$ (3.4 $\mu$m) and {\it WISE} $W2$ (4.6 $\mu$m) where the photometric error ($\sigma \approx 0.05$\,mag) appears to be comparable or larger than the intrinsic scatter. This suggests that additional observations at these wavelengths could improve the inferred distances to these sources further. With $\sim100,000$ RR Lyrae stars expected throughout the Galaxy, the precision dust extinction measurements towards 134 lines-of-sight offer a proof of concept for using such sources to make 3D tomographic maps of dust throughout the Milky Way. We find a small but significant increase (3 per cent) in the effective extinction towards sources far from the Galactic plane relative to the expectation from recent dust maps and we suggest several explanations. As an application of the methodology, we infer the distance to the RRc-type star RZCep at low Galactic latitude ($b = 5.5^\circ$) to be $\mu = 8.0397 \pm 0.0123$\,mag (405.4 $\pm$ 2.3\,pc) with colour excess $E(B - V) = 0.2461 \pm 0.0089$\,mag. This distance, equivalent to a parallax of $2467 \pm 14$ microarcsec, is consistent with the published HST parallax measurement but with an uncertainty that is 13 times smaller than the HST measurement. If our measurements (and methodology) hold up to scrutiny, the distances to these stars have been determined to an accuracy comparable to those expected with Gaia. As RR Lyrae are one of the primary components of the cosmic distance ladder, the achievement of sub-1 per cent distance errors within a formalism that accounts for dust extinction may be considered a strong buttressing of the path to eventual 1 per cent uncertainties in Hubble's constant.
\end{abstract}


\section{Introduction}

RR Lyrae stars are old (age $\gtrsim10\times10^9$ yr) Population II pulsating stars that exist throughout the Milky Way Bulge, Disc, and Halo. At optical wavebands they are variable with peak-to-peak amplitudes up to about 1 mag. This amplitude generally diminishes with increasing  wavelength to around 0.3 mag in the mid-infrared ($\lambda \sim 4 \mu$m). The heat generation and gravitational support of RR Lyrae stars comes from the fusion of helium in the core and hydrogen in a shell surrounding the core. RR Lyrae stars have specific values of temperature, luminosity, and radius such that they exist in the instability strip of the Hertzsprung--Russell diagram. In this slice of stellar parameter-space stars are unstable to radial oscillation, and RR Lyrae stars oscillate with periods ranging from about 0.2 to 0.9 d.\footnote{On weeks-long periods, some stars exhibit peak-amplitude variations called the Blazhko effect (cf.~subsection\,\ref{PM_relation_plot} for an additional description).} This, and oscillating temperature changes, leads to periodic luminosity variability. \citet{1964ARA&amp;A...2...23P} and \citet{1995CAS....27.....S} both provide excellent reviews of RR Lyrae pulsating variable stars. 

RR Lyrae (and other pulsational variables, namely Cepheids) have inspired more than a century's worth of close attention because of the correlation of their fundamental oscillation period and luminosity. This empirical relation, supported by theoretical modelling \citepeg{2004ApJS..154..633C}, has led to their use as primary distance indicators within the Milky Way and to the nearest neighbouring galaxies. With an effective range of ${\sim100 {\rm ~kpc}}$ (for a limiting AB mag of ${\sim20}$), they bridge the gap in the cosmic distance ladder between trigonometric parallax and the Tip of the Red Giant Branch (TRGB) method. RR Lyrae stars can also serve to calibrate Cepheid distances, most significantly by precisely measuring the distance and morphology of the Magellanic Clouds.

Distance determinations are achieved by leveraging empirical RR Lyrae period--magnitude relations to infer an RR Lyrae star's intrinsic luminosity in a given waveband (absolute magnitude, $M$) from its measured oscillation period. In prior work this relation is commonly called the ``period--luminosity'' relation, but here we prefer to use the term ``period--magnitude'' relation to distinguish that it is not a predictor of bolometric luminosity, but instead of absolute magnitude in a given waveband.

With absolute magnitude for a given waveband in hand, the distance modulus, $\mu$, is then calculated after observing the mean-flux apparent magnitude of the star in the same waveband, $m$,\footnote{Mean-flux magnitude is derived from a star's phase-folded light curve; the specifics are described in Section\,\ref{methods}.} and using the colour excess, $E\left(B-V\right)$, to correct for extinction due to light scattering by interstellar dust grains. For waveband $j$, the equation for distance modulus, which is common for all bands, is
\begin{equation}
\label{eqn:distance_modulus}
\mu = m_j - M_j - E\left(B-V\right) \times \left( a_j R_V  + b_j \right),
\end{equation}
where $a_j$ and $b_j$ are the wavelength-specific parameters for the interstellar extinction law defined in \cite{1989ApJ...345..245C} and $R_V$ is the extinction law factor (${R_V = A_V / E[B-V]}$) with value equal to 3.1 for the diffuse interstellar medium adopted from \cite{1975A&A....43..133S}.

The period--magnitude relations of RR Lyrae stars have previously been primarily studied in the near-infrared $K$-band \citep{2006MNRAS.372.1675S} and mid-infrared bands of the {\it Wide-field Infrared Survey Explorer} ({\it WISE}) all-sky satellite survey mission \citep{2011ApJ...738..185K,2013ApJ...776..135M,2014MNRAS.439.3765D,2014MNRAS.440L..96K}. The slope of the period--magnitude relation at the shorter-wavelength optical bands is shallower, and a stronger correlation has been found between metallicity, [Fe/H], and optical (generally $V$-band) absolute magnitude. However, shorter-wavelength observations exhibit large scatter about a linear relation and there is no clear consensus on the necessity of the inclusion of secondary and/or nonlinear terms. The interested reader is referred to Section 5 of \citet{2006ARA&A..44...93S} for a review of optical ($V$-band) implementations of RR Lyrae stars as distance indicators.

In this analysis for the first time empirical RR Lyrae period--magnitude relations are simultaneously derived for 13 wavebands between the ultraviolet and mid-infrared. The calibration dataset is comprised of 134 RR Lyrae stars with photometry data combined from four astronomical observing facilities (two ground-based telescopes and the space-based {\it Hipparcos} and {\it WISE} satellites). Distances for the calibration RR Lyrae stars are determined with median fractional error of 0.66 per cent, and the multi-wavelength data are also used to solve for the colour excess to each calibration star.

The improved waveband-specific period--magnitude relations presented here, as well as the Bayesian methodology for simultaneously calibrating or applying any subset of the 13 relations, represent a significant advancement in the use of RR Lyrae stars to measure distance. The claimed level of precision compares to (and even rivals) the expected astrometric precision of Gaia, the space-based, parallax/astrometry mission launched by the European Space Agency in December 2013 \citep{gaiabrochure}.

This paper is outlined as follows. We present a description of the  ground-based optical, {\it Hipparcos}, ground-based near-infrared, and {\it WISE} datasets in Section\,\ref{data_desc}. In Section\,\ref{methods} we review our light curve analysis methodology, describing how mean-flux magnitudes are measured for the sample. In Section\,\ref{derivations} we present our Bayesian simultaneous linear regression formalism, extended from our prior work, and the resultant period--magnitude relations. In Section\,\ref{discussion} we relate additional findings of note resulting from the period--magnitude relation fits. In Section\,\ref{application} we demonstrate how the multi-band period--magnitude relations are applied to estimate the distance to RZCep, which had been excluded from the period--magnitude relation fits owing to high, and poorly constrained, interstellar extinction. Finally, in Section\,\ref{conclusions} we discuss the conclusions and future implications of this work.

\section[]{Data Description}\label{data_desc}

The RR Lyrae calibration sample used in this work is based upon the catalogue of 144 relatively local ($\leqslant 2.5$ kpc) RR Lyrae variables developed by \cite{1998AA...330..515F}. Six of these stars are excluded from our present analysis because of minimal light curve data and poor harmonic model fits determined via the procedure described in Section\,\ref{methods}. Another three stars (ARPer, RZCep, and BNVul) are excluded because of their large and poorly constrained colour excess values (these stars lie too close to the Galactic Plane for the \citealt{1998ApJ...500..525S} and \citealt{2011ApJ...737..103S} dust maps to provide accurate colour excess measurements). And, one additional star, ATSer, was excluded because only {\it Hipparcos} and $W3$ photometry was available for this star, and the $W3$ magnitude was a significant outlier in prior period--magnitude relation fits.

In total, the calibration sample is 134 stars, with 637 band-specific light curves composed of 33,630 epochs. Table\,\ref{tab:catalogue} provides complete observable prior and fitted posterior data for the calibration sample.

\begin{table*}
 \centering
\begin{minipage}{6.70 in}
  \caption{First few rows of the catalogue of calibration RR Lyrae stars (transposed into columns to fit on the page). Observed values, measured priors, and inferred posterior variables are shown. The full table can be accessed as a text file in the publication's online data.}
  \begin{tabular}{c | rrrrrr}
  \hline
  \cutinhead{Measurements and Priors}{7}
Name & \multicolumn{1}{c}{AACMi} & \multicolumn{1}{c}{ABUMa} & \multicolumn{1}{c}{AEBoo} & \multicolumn{1}{c}{AFVel} & \multicolumn{1}{c}{AFVir} & ..... \\
Type & \multicolumn{1}{c}{RRab} & \multicolumn{1}{c}{RRab} & \multicolumn{1}{c}{RRc} & \multicolumn{1}{c}{RRab} & \multicolumn{1}{c}{RRab} & ..... \\
Blazhko Affected? & \multicolumn{1}{c}{False} & \multicolumn{1}{c}{False} & \multicolumn{1}{c}{False} & \multicolumn{1}{c}{True} & \multicolumn{1}{c}{False} & ..... \\
Period (d) & \multicolumn{1}{c}{0.4763} & \multicolumn{1}{c}{0.5996} & \multicolumn{1}{c}{0.3149} & \multicolumn{1}{c}{0.5274} & \multicolumn{1}{c}{0.4837} & ..... \\
$[{\rm Fe}/{\rm H}]$ & \multicolumn{1}{c}{$-0.15$} & \multicolumn{1}{c}{$-0.49$} & \multicolumn{1}{c}{$-1.39$} & \multicolumn{1}{c}{$-1.49$} & \multicolumn{1}{c}{$-1.33$} & ..... \\
$\mu_{\rm Prior}$ & $10.4730 \pm 0.1384$ & $10.0551 \pm 0.1326$ & $9.9779 \pm 0.1252$ & $10.2164 \pm 0.1282$ & $11.1113 \pm 0.1258$ & ..... \\
$E(B-V)_{\rm SF}$ & $0.0997 \pm 0.0029$ & $0.0226 \pm 0.0012$ & $0.0230 \pm 0.0012$ & $0.2783 \pm 0.0068$ & $0.0208 \pm 0.0010$ & ..... \\
$m_U$ & \multicolumn{1}{c}{\nodata} & $11.6575 \pm 0.0026$ & \multicolumn{1}{c}{\nodata} & \multicolumn{1}{c}{\nodata} & \multicolumn{1}{c}{\nodata} & ..... \\
$m_B$ & \multicolumn{1}{c}{\nodata} & $11.3352 \pm 0.0017$ & \multicolumn{1}{c}{\nodata} & \multicolumn{1}{c}{\nodata} & \multicolumn{1}{c}{\nodata} & ..... \\
$m_{hipp}$ & $11.6734 \pm 0.0059$ & $11.0349 \pm 0.0038$ & $10.7220 \pm 0.0034$ & $11.5419 \pm 0.0059$ & $11.8920 \pm 0.0104$ & ..... \\
$m_V$ & \multicolumn{1}{c}{\nodata} & $10.8694 \pm 0.0016$ & \multicolumn{1}{c}{\nodata} & \multicolumn{1}{c}{\nodata} & \multicolumn{1}{c}{\nodata} & ..... \\
$m_R$ & \multicolumn{1}{c}{\nodata} & $10.5775 \pm 0.0018$ & \multicolumn{1}{c}{\nodata} & \multicolumn{1}{c}{\nodata} & \multicolumn{1}{c}{\nodata} & ..... \\
$m_I$ & \multicolumn{1}{c}{\nodata} & $10.3370 \pm 0.0075$ & \multicolumn{1}{c}{\nodata} & \multicolumn{1}{c}{\nodata} & \multicolumn{1}{c}{\nodata} & ..... \\
$m_z$ & \multicolumn{1}{c}{\nodata} & $10.7343 \pm 0.0066$ & \multicolumn{1}{c}{\nodata} & \multicolumn{1}{c}{\nodata} & \multicolumn{1}{c}{\nodata} & ..... \\
$m_J$ & \multicolumn{1}{c}{\nodata} & $10.0556 \pm 0.0065$ & \multicolumn{1}{c}{\nodata} & \multicolumn{1}{c}{\nodata} & \multicolumn{1}{c}{\nodata} & ..... \\
$m_H$ & \multicolumn{1}{c}{\nodata} & $9.7284 \pm 0.0059$ & \multicolumn{1}{c}{\nodata} & \multicolumn{1}{c}{\nodata} & \multicolumn{1}{c}{\nodata} & ..... \\
$m_K$ & \multicolumn{1}{c}{\nodata} & $9.7181 \pm 0.0072$ & \multicolumn{1}{c}{\nodata} & \multicolumn{1}{c}{\nodata} & \multicolumn{1}{c}{\nodata} & ..... \\
$m_{W1}$ & $10.2375 \pm 0.0057$ & $9.5697 \pm 0.0047$ & $9.7129 \pm 0.0042$ & $9.9828 \pm 0.0067$ & $10.6980 \pm 0.0062$ & ..... \\
$m_{W2}$ & $10.2493 \pm 0.0053$ & $9.5920 \pm 0.0043$ & $9.7212 \pm 0.0041$ & $9.9809 \pm 0.0114$ & $10.7121 \pm 0.0085$ & ..... \\
$m_{W3}$ & $10.2216 \pm 0.0539$ & $9.5304 \pm 0.0259$ & $9.6652 \pm 0.0235$ & $9.9081 \pm 0.0361$ & \multicolumn{1}{c}{\nodata} & ..... \\
\cutinhead{Posterior Inferences}{7}
$\mu_{\rm Post}$ & $10.5864 \pm 0.0135$ & $10.1514 \pm 0.0146$ & $9.9580 \pm 0.0167$ & $10.4202 \pm 0.0132$ & $11.0736 \pm 0.0136$ & ..... \\
$E(B-V)_{\rm Post}$ & $0.1469 \pm 0.0157$ & $0.0955 \pm 0.0088$ & $0.0456 \pm 0.0157$ & $0.1656 \pm 0.0159$ & $0.0682 \pm 0.0158$ & ..... \\
$M_U$ & \multicolumn{1}{c}{\nodata} & $1.0458 \pm 0.0433$ & \multicolumn{1}{c}{\nodata} & \multicolumn{1}{c}{\nodata} & \multicolumn{1}{c}{\nodata} & ..... \\
$M_B$ & \multicolumn{1}{c}{\nodata} & $0.7872 \pm 0.0377$ & \multicolumn{1}{c}{\nodata} & \multicolumn{1}{c}{\nodata} & \multicolumn{1}{c}{\nodata} & ..... \\
$M_{hipp}$ & $0.5973 \pm 0.0517$ & $0.5649 \pm 0.0314$ & $0.6120 \pm 0.0522$ & $0.5697 \pm 0.0517$ & $0.5908 \pm 0.0522$ & ..... \\
$M_V$ & \multicolumn{1}{c}{\nodata} & $0.4189 \pm 0.0297$ & \multicolumn{1}{c}{\nodata} & \multicolumn{1}{c}{\nodata} & \multicolumn{1}{c}{\nodata} & ..... \\
$M_R$ & \multicolumn{1}{c}{\nodata} & $0.1768 \pm 0.0258$ & \multicolumn{1}{c}{\nodata} & \multicolumn{1}{c}{\nodata} & \multicolumn{1}{c}{\nodata} & ..... \\
$M_I$ & \multicolumn{1}{c}{\nodata} & $0.0077 \pm 0.0220$ & \multicolumn{1}{c}{\nodata} & \multicolumn{1}{c}{\nodata} & \multicolumn{1}{c}{\nodata} & ..... \\
$M_z$ & \multicolumn{1}{c}{\nodata} & $0.4383 \pm 0.0197$ & \multicolumn{1}{c}{\nodata} & \multicolumn{1}{c}{\nodata} & \multicolumn{1}{c}{\nodata} & ..... \\
$M_J$ & \multicolumn{1}{c}{\nodata} & $-0.1827 \pm 0.0171$ & \multicolumn{1}{c}{\nodata} & \multicolumn{1}{c}{\nodata} & \multicolumn{1}{c}{\nodata} & ..... \\
$M_H$ & \multicolumn{1}{c}{\nodata} & $-0.4775 \pm 0.0160$ & \multicolumn{1}{c}{\nodata} & \multicolumn{1}{c}{\nodata} & \multicolumn{1}{c}{\nodata} & ..... \\
$M_K$ & \multicolumn{1}{c}{\nodata} & $-0.4672 \pm 0.0163$ & \multicolumn{1}{c}{\nodata} & \multicolumn{1}{c}{\nodata} & \multicolumn{1}{c}{\nodata} & ..... \\
$M_{W1}$ & $-0.3746 \pm 0.0144$ & $-0.5984 \pm 0.0153$ & $-0.2530 \pm 0.0170$ & $-0.4663 \pm 0.0145$ & $-0.3875 \pm 0.0146$ & ..... \\
$M_{W2}$ & $-0.3529 \pm 0.0143$ & $-0.5697 \pm 0.0151$ & $-0.2418 \pm 0.0170$ & $-0.4570 \pm 0.0172$ & $-0.3688 \pm 0.0158$ & ..... \\
$M_{W3}$ & $-0.3681 \pm 0.0555$ & $-0.6232 \pm 0.0297$ & $-0.2938 \pm 0.0288$ & $-0.5158 \pm 0.0385$ & \multicolumn{1}{c}{\nodata} & ..... \\

\hline
\label{tab:catalogue}
\end{tabular}
\end{minipage}
\end{table*}

The calibration sample contains RR Lyrae stars belonging to both of the two most common subtypes: RRab (115) and RRc (19). RRab stars oscillate at their fundamental period, $P_f$, and RRc stars oscillate at their first overtone period, $P_{fo}$. The RRc stars' periods must be ``fundamentalised'' before deriving the period--magnitude relations. As in \cite{2004ApJ...610..269D}, an RRc star's fundamentalised period is given by 
\begin{equation}
\label{eqn:period_fundamentalize}
\log_{10}\left(P_{f}\right) = \log_{10}\left(P_{fo}\right) + 0.127.
\end{equation}

\subsection[]{Colour excess and distance modulus priors}\label{priors_subsection}

Line-of-sight ${E(B-V)}$ colour excess values published in \citet{1998ApJ...500..525S} and \citet{2011ApJ...737..103S} were retrieved from the NASA/IPAC Infrared Science Archive. These colour excess values estimate the total cumulative interstellar extinction due to dust. In practice, the calibration stars are embedded in the Galaxy and the dust maps, which were derived from far-infrared imaging, are averaged over large (tens of arcminute) scales. The former means that the true colour excess can be significantly less than the published value for that line of sight (even approaching zero if the star is close enough) and the latter implies that the published values should be considered to have significantly larger uncertainty bounds when applied to precise lines of sight terminating at unresolved point sources. 

In order to begin the Markov Chain Monte Carlo (MCMC) regression traces (Section\,\ref{sec:mcmc}) and ultimately fit for colour excess posteriors, the \cite{2011ApJ...737..103S} values were adapted into prior colour excess distributions according to the following procedure. If ${E(B-V)_{\rm SF} > 0.125}$, then the prior distribution was set to be uniform ${\mathcal{U}(0, 2.5 \times E(B-V)_{\rm SF})}$. Otherwise, if ${E(B-V)_{\rm SF} \le 0.125}$, then the prior distribution was set to be ${\mathcal{U}(0, 0.125)}$.

Prior distributions for the calibrator distance moduli were derived as in \cite{2011ApJ...738..185K} and \cite{2014MNRAS.440L..96K}. {\it Hipparcos} photometry \citep{1997ESASP1200.....P} were transformed into $V$-band \citep{1998ApJ...508..844G}, corrected for dust extinction (using the line-of-sight extinction from \citealt{2011ApJ...737..103S} and the $R$ factor from \citealt{1975A&A....43..133S}), and combined with the \cite{postHipp..book} $M_V$--[Fe/H] relation to yield prior distance moduli, $\mu_{\rm Prior}$. Precise trigonometric parallax angles for four of the stars (RRLyr, UVOct, XZCyg, and SUDra; all of the RRab subclass) have been previously measured with the {\it Hubble Space Telescope} (HST) and published \citep{2011AJ....142..187B}.\footnote{The RRc star RZCep also has an HST-measured parallax, but this star was rejected from our fit because of low galactic latitude and, consequently, a poorly constrained prior colour excess value.} For these four stars the more precise, parallax-derived distance moduli were used in the period--magnitude relation fits. We note that the the distance moduli derived from the metallicity--magnitude relation for these four stars is in statistical agreement (within $2\sigma$) with the parallax-derived distances.

\subsection[]{\textbfit{Hipparcos} photometry}

The European Space Agency {\it Hipparcos} astrometry satellite was launched in August 1989 and operated until March 1993, ultimately producing a catalogue of photometry, parallax, and, in the case of variable stars, light curves, published in \cite{1997ESASP1200.....P}. {\it Hipparcos} obtained light curves for 186 RR Lyrae stars, 134 of which serve as the calibration sample for the period--magnitude relations derived in this work.

Since {\it Hipparcos} was primarily an astrometry mission, its imaging detector used a broadband visible light passband, defined primarily by the response function of the detector, an unfiltered S20 image dissector scanner. \citet{2000PASP..112..961B} characterizes the {\it Hipparcos} waveband, commonly referred to as $H_P$. Throughout this work, to reduce potential confusion with the near-infrared $H$-band, the {\it Hipparcos} waveband is referred to as $hipp$. The effective wavelength of the $hipp$ waveband is taken to be 0.517 $\mu$m, and the bandpass itself is substantially broader than $V$-band (see Fig.\,2 of \citealt{2000PASP..112..961B}).

{\it Hipparcos} was a temporally dense all-sky survey, and thus it provides the most complete and numerous light curve data for the RR Lyrae calibration sample. All 134 calibrator stars have $hipp$ light curves, which are composed of 11,822 epochs.

\subsection[]{Optical photometry}

Ground-based optical light curves were obtained with the Nickel 1-m telescope and Direct Imaging Camera at Lick Observatory in California. Imaging data was collected in the $U$, $B$, $V$, $R$, $I$, and Sloan Digital Sky Survey (SDSS) $z$ wavebands during 26 nights between 2010 May 4 and 2013 February 4. Standard image reduction was conducted using common Python scientific computing modules [using PyFITS \citep{1999ASPC..172..483B} for image reading and writing] and aperture photometry was measured with SExtractor \citep{1996A&AS..117..393B}. Photometric calibration was performed using observations of Landolt standards in the $U$, $B$, $V$, $R$, and $I$ wavebands (\citealt{1992AJ....104..340L}, updated by \citealt{2009AJ....137.4186L}), and SDSS standards for the $z$ waveband \citep{2002AJ....123.2121S}.

The Direct Imaging Camera filter wheel could only accommodate four filters at one time, and so preference was given to $U$, $B$, $V$, and $R$ for the first 21 nights (before 2012). The $I$ and $z$ filters replaced the $U$ and $B$ filters in the 5 observing nights after and including 2012 November 6. The targets for these last 5 nights were repeats of stars already observed during the first 21 nights, and the primary purpose was to supplement the calibration waveband coverage of the sample.

In the $U$-band 22 light curves were obtained, consisting of 1409 epochs. 
In the $B$-band 24 light curves were obtained, consisting of 1599 epochs. 
In the $V$-band 25 light curves were obtained, consisting of 1991 epochs. 
In the $R$-band 25 light curves were obtained, consisting of 2031 epochs. 
In the $I$-band 9 light curves were obtained, consisting of 410 epochs. 
And, in the $z$-band 9 light curves were obtained, consisting of 400 epochs. 

\subsection[]{Near-infrared photometry}

Observations in the $J$, $H$, and $K_{\rm short}$ (herein abbreviated simply as $K$) wavebands were conducted between 2009 April 14 and 2011 May 18 with the 1.3-m Peters Automated Infrared Telescope ({\it PAIRITEL}; \citealt{2006ASPC..351..751B}) at Fred Lawrence Whipple Observatory in Arizona. {\it PAIRITEL} was the robotized 2MASS North telescope mated with the repurposed 2MASS South camera. As such, the near-infrared wavebands used in the present work are identical to the 2MASS photometric system, and photometric calibration was conducted using reference stars contained within the same ${8.53^\prime \times 8.53^\prime}$ field of view. The near-infrared images were reduced and coadded with the software pipeline described in the following subsection. Aperture photometry was measured with SExtractor.

In the $J$-band 18 light curves were obtained, consisting of 1293 epochs. 
In the $H$-band 17 light curves were obtained, consisting of 1247 epochs. 
And, in the $K$-band 22 light curves were obtained, consisting of 1512 epochs. 

\subsubsection[]{{\it PAIRITEL} reduction pipeline}

Because {\it PAIRITEL} reused the 2MASS camera and unaltered readout electronics, each epoch consisted of multiple exposure triplets separated by $\sim$dozen seconds during which a small dither offset was enforced. Each single exposure in the triplet had an exposure time of 7.8 s. A single epoch generally consisted of 8 or 9 triplets, making for a total integration time of $\sim3$ to $\sim3.5$ minutes.

In support of this work on RR Lyrae period--magnitude relations, as well as the prime science goal of {\it PAIRITEL} to followup gamma-ray burst (GRB) afterglows, a new image reduction and co-addition pipeline was developed for the robotic telescope and deployed for near-real time operation. This software was the third and final reduction pipeline developed for {\it PAIRITEL}. It operated autonomously in concert with the telescope as new data was gathered each night, often providing reduced and coadded images within a few minutes of the end of an observation. This was particularly beneficial for quickly reacting to GRBs and issuing GCN circulars. The reduction pipeline also provided invaluable near-real time diagnostic information for the telescope supervisors when troubleshooting mechanical, technical, or telescope control system-related faults.

The 2MASS camera uses two dichroics and three near-infrared detectors to simultaneously record the $J$, $H$, and $K$ exposures. For the most part, the reduction pipeline operates on each waveband independently. However, because the images are taken simultaneously in each band, the relative and absolute astrometric solutions for the images need only be solved for $J$ and can then be applied to the two longer-wavelength (and less sensitive) $H$ and $K$ exposures.

The constrained image readout mode of {\it PAIRITEL} dictated much of how the reduction pipeline operated. Each ${7.8~{\rm s}}$ integration of a triplet exposure (called a ``long read'') was preceded by a ``short read'' of ${0.051~{\rm s}}$. The short read served as a bias read for the long read, and was subtracted from the long read as the first step in the reduction process. The short reads themselves were also processed to produce final coadded images with very short total exposure times. The advantage of processing the short reads is recovery of extremely bright sources that otherwise saturate in the long reads. This was the intended avenue for photometering the nearby bright RR Lyrae stars, such as RRLyr itself, but ultimately the photometric precision recoverable from the reduced and coadded short reads was found to be unacceptable.

In the near-infrared, the brightness of the atmosphere is significant and must be subtracted to improve the signal-to-noise ratio of astrophysical sources. The reduction pipeline creates median sky background images by masking pixels suspected to fall on sources and stacking temporally-adjacent images. The sky brightness fluctuates on 5- to 10-minute timescales, so for a given ``target'' exposure the pipeline uses the images recorded within $\pm5$ minutes to create this median sky flux image. (Of course, if the target exposure is within 5 minutes of the beginning or end of the observation period, then fewer adjacent images contribute to its sky flux image.)

It was found that the detector response varied significantly, and in a correlated manner, with the read-cycle position of the long reads in the triplet exposures. To account for this, a different sky flux image is produced for each of the three long reads in a triplet exposure, wherein only the first long read of each contributing triplet exposure is combined into the sky flux image corresponding to the first long read of the target triplet exposure, and so on for the second and third reads in the cycle.

The accuracy of this sky brightness subtraction procedure relies heavily upon correctly masking pixels containing flux from astrophysical sources from contributing to the median sky flux image. The reduction pipeline runs this sky subtraction procedure twice, first with a preliminary source pixel mask and then with a more refined, and conservative, source pixel mask constructed from the images resulting from summing long reads of each triplet exposure after subtracting the first iteration of the sky flux images. The source pixel masks were generated by employing a median absolute deviation outlier detection algorithm in combination with the objects check image output from SExtractor. The raw source pixel masks were then Gaussian smoothed (blurred) to expand the masked pixel area and account for diffuse emission from extended sources and the telescope's PSF. The dither steps between each triplet exposure were large enough to ``step over'' the footprints of unsaturated (and most saturated) point sources, as well as most galaxies with radii $\lesssim 30^{\prime\prime}$. 

After the sky flux subtraction, each triplet exposure is directly pixel-wise summed to create a ``triplestack''. Each pixel is $2^{\prime\prime} \times 2^{\prime\prime}$, and the telescope jitter was far smaller, so this does not result in any significant smearing. The final step in the reduction process is to coadd the images and produce mosaics, but before this can be done the relative dither offsets must be measured from the pixel data and written into the FITS header WCS keywords. Note that an absolute astrometric solution is not necessary at this step, only a WCS solution that incorporates precisely correct relative offsets between the triplestacks. To accomplish this, the reduction pipeline runs SExtractor on each $J$-band triplestack and analyes the resultant catalogs to identify the deepest triplestack image. This deep triplestack, generally the image with the most well-detected sources, serves as the reference image from which the pixel offsets of the other images in the sequence are measured. The relative sky position offsets and rotations between the $J$, $H$, and $K$ detectors are well known and constant, so it is only necessary to measure the offsets in the $J$-band triplestack sequence.

A normalized cross-correlation image-alignment program (specially developed by E.~Rosten) is used to measure the pixel offsets between the reference triplestack and all other images in the sequence. In addition to the image pair, the alignment program also requires an approximate pixel offset (derived from the telescope control system's imprecise pointing data) and a search box width. The computed pixel offsets are accurate at the sub-pixel level.

With relative pixel offsets in hand, the reduction pipeline writes appropriate WCS information into the FITS headers of the $J$, $H$, and $K$ triplestack sequences and then uses Swarp \citep{2002ASPC..281..228B} to median-combine and mosaic the reduced imaging data. In the mosaicing process the pixel resolution is changed from 2$^{\prime\prime}$ to 1$^{\prime\prime}$. The final astrometry is solved using {\it Astrometry.net} \citep{2010AJ....139.1782L}, although sometimes the pipeline falls back on Scamp \citep{2006ASPC..351..112B} and then, if Scamp also fails, a specifically-developed pattern-matching Python program is employed.

\subsection[]{\textbfit{WISE} photometry}

Mid-infrared light curve photometry data were obtained from the AllWISE Data Release of the {\it Wide-field Infrared Survey Explorer} ({\it WISE}\,) and its extended NEOWISE mission \citep{2010AJ....140.1868W, 2011ApJ...731...53M}. {\it WISE} provides imaging data in four mid-infrared wavebands: $W1$ centred at 3.4\,$\mu$m, $W2$ centred at 4.6\,$\mu$m, $W3$ centred at 12\,$\mu$m, and $W4$ centred at 22\,$\mu$m. Although the original {\it WISE} mission was designed for static science goals, the orbit and survey strategy of the {\it WISE} spacecraft (described in \citealt{2010AJ....140.1868W}) are highly conducive to recovering light curves of periodic variables with periods $\lesssim 1.5$\,d, which is well-matched to RR Lyrae variables.

The AllWISE Data Release (made public 2013 November 13) combines the 4-Band Cryogenic Survey (main {\it WISE} mission covering the full sky 1.2 times from 2010 January 7 to 2010 August 6), the 3-Band Cryogenic survey (first three wavebands, 30 per cent of the sky from 2010 August 6 to 2010 September 29), and the NEOWISE post-cryogenic survey (first two wavebands, covering 70 per cent of the sky from 2010 September 20 to 2011 February 1). The individual photometry epochs were retrieved from the AllWISE Multiepoch Photometry Database.

{\it WISE}, like {\it Hipparcos}, was an all-sky survey and thus the AllWISE Data Release provides very good coverage of the calibration sample. In the $W1$-band 126 light curves were obtained, consisting of 4202 epochs. In the $W2$-band 127 light curves were obtained, consisting of 4204 epochs. And, in the $W3$-band 79 light curves were obtained, consisting of 1510 epochs. Significantly fewer stars were detected and provided light curves accepted into the calibration sample in $W3$ because the $W3$ detector was not as sensitive and was not operating for the NEOWISE period. Additionally, all $W4$ data are rejected from the present work because only the few brightest calibration RR Lyrae stars were detected in that bandpass.

\section{Light Curve Analysis Methods}\label{methods}

The light curve analysis methods employed in this work are an evolution of those described in \cite{2011ApJ...738..185K} and \cite{2014MNRAS.440L..96K}. Each band-specific light curve is parametrically resampled (assuming a normal distribution) 500 times to fit 500 harmonic models using the adopted pulsation period from \cite{1998AA...330..515F}. Thus, 500 realizations of the mean-flux magnitude are measured, and the standard deviation of this distribution is taken to be the uncertainty on the mean-flux magnitude. These are the observed mean-flux magnitudes reported in Table\,\ref{tab:catalogue} and are not corrected for interstellar extinction.

\begin{figure}
	\centering
	\includegraphics{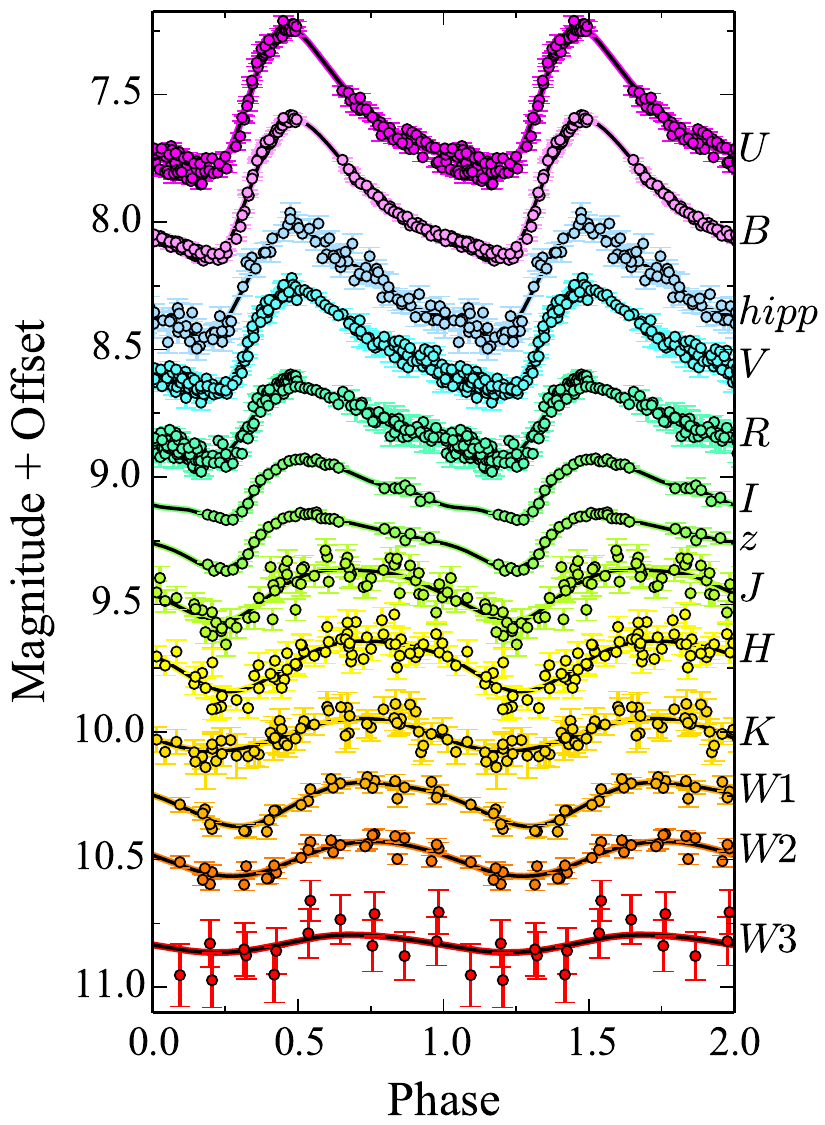}
	\caption{Light curves of RR Lyrae star ABUMa with period 0.6 d. The black solid curves are the mean of 500 bootstrapped harmonic models fitted to the light curve data in each waveband. The phase migration of the light curve peak brightness is real. An arbitrary relative phase offset was applied to the {\it Hipparcos} light curve because those data were observed  between 1989 and 1993 (whereas the rest of the data were acquired between 2009 and 2013). Nearly 10,000 cycles occurred between the last {\it Hipparcos} observation and the first {\it PAIRITEL} observation, and this expanse was too large to accurately phase match.}
	\label{fig:ABUMa_light_curve}
\end{figure}

The 500 harmonic models generated by the bootstrapping procedure were averaged to produce a mean harmonic model. Fig.\,\ref{fig:ABUMa_light_curve} shows the phase-folded light curve data and mean harmonic models for ABUMa (which was specifically selected to show a well-observed RR Lyrae calibration star with complete 13-waveband data). 

The mean harmonic model yields a robust light curve amplitude. Furthermore, the standard deviation of the 500 harmonic models at each phase value provides a metric of how well the shape of the true light curve is recovered in the photometry data (if there is a lot of spread in the distribution of harmonic models, then the photometry is not accurate enough to reveal the shape of the true brightness oscillation). To improve the quality of the dataset used in the period--magnitude relation fits, any light curve with a bootstrapped harmonic model maximum standard deviation larger than its robust amplitude measurement was excluded. This procedure serves to ensure that only stars with light curves well-fit by the harmonic model (i.e., those exhibiting clear sinusoidal-like oscillation) are used in the period--magnitude relation fits.

The summary information given above in Section\,\ref{data_desc} for the number of light curves obtained in each waveband has already taken into account the results of this quality selection process. For example, most of the diminution in the number of $W3$ light curves (79) as compared to $W1$ (126) or $W2$ (127) is due to this requirement that the model uncertainty be less than the light curve amplitude.

\section[]{Period--Magnitude Relations}\label{derivations}

The present derivation of period--magnitude relations is similar to the Bayesian approach first described in \cite{2011ApJ...738..185K} and later formalised in \cite{2012Ap&amp;SS.341...83K}. A significant advancement over previous implementations is the inclusion of colour excess as a model parameter. Our statistical model of the period--magnitude relationship is 
\begin{equation}
\label{eqn:general_PLR}
m_{ij}=\mu_i + M_{0,j} + \alpha_j \log_{10}\left(P_i/P_0\right) + E(B-V)_i\left( a_j R_V + b_j \right) + \epsilon_{ij},
\end{equation}
where $m_{ij}$ is the observed apparent magnitude of the $i$th RR Lyrae star in the $j$th waveband, $\mu_i$ is the distance modulus for the $i$th RR Lyrae star, $M_{0,j}$ is the absolute magnitude zero point for the $j$th waveband, $\alpha_j$ is the slope in the $j$th waveband, $P_i$ is the fundamentalised period of the $i$th RR Lyrae star in days, $P_0$ is a period normalisation factor (we use the mean fundamentalised period of the calibration sample, ${P_0 = 0.52854~{\rm d}}$), ${E(B-V)_i}$ is the colour excess of the $i$th RR Lyrae star, $a_j$ and $b_j$ are the wavelength-specific parameters for the interstellar extinction law defined in \cite{1989ApJ...345..245C}, $R_V$ is the extinction law factor (${R_V = A_V / E[B-V]}$) with value equal to 3.1 for the diffuse interstellar medium adopted from \cite{1975A&A....43..133S}, and the $\epsilon_{ij}$ error terms are independent zero-mean Gaussian random deviates with variance ${(\sigma_{{\rm intrinsic},j}^2 +  \sigma_{m_{ij}}^2)}$. We note that the $\epsilon_{ij}$ error terms are defined differently than in previous work to allow for the model to fit wavelength-dependent intrinsic period--magnitude relation scatter ($\sigma_{{\rm intrinsic},j}$). This additive error term, which we call the intrinsic scatter, describes the residual about the best-fit period--magnitude relation in each waveband which cannot be accounted for by instrumental photometric error. Such scatter would naturally be expected if there are unmodelled wavelength-sensitive dependencies (such as with metallicity) on the period--magnitude relation.

To perform the Bayesian regression a design matrix ${\mathbf X}$ is constructed for the model expressed in Equation\,\ref{eqn:general_PLR}. ${\mathbf X}$ has dimensions $637\times294$. Each of the 637 light curves produced one mean-flux magnitude measurement which is represented by a row in ${\mathbf X}$. The terms in Equation\,\ref{eqn:general_PLR} with $i$-dependence ($\mu_i$ and ${E[B-V]_i}$, where each RR Lyrae star is fit with one value) each require 134 columns. And, the terms in Equation\,\ref{eqn:general_PLR} with $j$-dependence ($M_{0,j}$ and $\alpha_j$, where each waveband is fit with one value) each require 13 columns.

We define the vector of model parameters, ${\mathbf b}$, which contains the 134 values of $\mu_i$, the 13 values of $M_{0,j}$, the 13 values of $\alpha_j$, and the 134 values of ${E(B-V)_i}$. The vector of observed mean-flux magnitudes, ${\mathbf m}_{\rm obs}$, is then given by the dot product of the design matrix and the vector of model parameters,
\begin{equation}
\label{eqn:model_dot_product}
{\mathbf m}_{\rm obs} = {\mathbf X} \cdot {\mathbf b}.
\end{equation}

The model parameters are fit by an implementation of MCMC sampling (\ref{sec:mcmc}) that iteratively refines the distributions of the model parameters until a converged steady-state is achieved. The fitting algorithm is run with the PyMC \citep{Patil:Huard:Fonnesbeck:2010:JSSOBK:v35i04} Python module, which leverages the distribution of the observed data vector ${\mathbf m}_{\rm obs}$ with variance given by ${(\sigma_{{\rm intrinsic},j}^2 +  \sigma_{m_{ij}}^2)}$, as well as the model parameter vector ${\mathbf b}$ and the associated variance on each model parameter. 

Initially ${\mathbf b}$ is populated with prior distributions and the MCMC sampling traces are run until convergence, after which 50,000 additional samples are drawn to record the fitted model parameter distributions (also called the posteriors). To avoid inappropriate biasing of the posterior distributions for the slope and intercepts, a wide normal distribution is adopted:
\begin{eqnarray}
\label{eqn:prior_normal_distributions}
M_{0,j, {\rm Prior}} &=& \mathcal{N}(0, 2^2), \\
\alpha_{j, {\rm Prior}} &=& \mathcal{N}(0, 5^2).
\end{eqnarray}
The prior distributions for distance modulus and colour excess are star-dependent and given in subsection\,\ref{priors_subsection}.

The summary results for the simultaneous 13-waveband period--magnitude relation fits are provided in Table\,\ref{tab:plr_parameters}. In the ensuing subsections more detail is provided for the execution of the MCMC fitting procedure, the posterior joint distributions for the 13 (zero point, intercept) pairs are illustrated and explained, and the comprehensive ${\log_{10}(P) - M}$ plot and a validation $\mu_{\rm Prior} - \mu_{\rm Post}$ plot are furnished.

\begin{table*}
 \centering
\begin{minipage}{4.45 in}
  \caption{Period--magnitude relation parameters and 1-$\sigma$ uncertainties. The band-specific form of the period--magnitude equation is $M = M_0 + \alpha  \log_{10} \left( P/P_0 \right)$, where $P_0=0.52854$ d. $\sigma_{\rm instrumental}$ is the average photometric uncertainty for the mean-flux magnitudes in each band and is dominated by the quality of the light curve data (although individual light curve consistency of each star does contribute). Note that $\sigma_{\rm instrumental}$ is not a model parameter, but is provided in the table for direct comparison with $\sigma_{\rm intrinsic}$, which is the fitted intrinsic scatter of the period--magnitude relation in each waveband. Fig.\,\ref{fig:sigmas} plots both $\sigma_{\rm intrinsic}$ and $\sigma_{\rm instrumental}$ as a function of wavelength.}
  \begin{tabular}{crrrr}
  \hline
band	 & \multicolumn{1}{c}{$M_0$ (intercept)} & \multicolumn{1}{c}{$\alpha$ (slope)} & \multicolumn{1}{c}{$\sigma_{\rm intrinsic}$} & \multicolumn{1}{c}{$\sigma_{\rm instrumental}$}\\ 
 \hline
$U$         & $0.9304\pm0.0584$     & $-0.3823\pm0.7130$    & $0.2358\pm0.0438$     & $0.0232\pm0.0175$  \\
$B$         & $0.7099\pm0.0237$     & $ 0.0129\pm0.3104$    & $0.0553\pm0.0126$     & $0.0145\pm0.0118$  \\
$hipp$      & $0.5726\pm0.0174$     & $-0.4625\pm0.2246$    & $0.0474\pm0.0079$     & $0.0098\pm0.0085$  \\
$V$         & $0.4319\pm0.0184$     & $-0.4091\pm0.2370$    & $0.0320\pm0.0079$     & $0.0106\pm0.0085$  \\
$R$         & $0.2638\pm0.0164$     & $-0.7461\pm0.2108$    & $0.0274\pm0.0072$     & $0.0091\pm0.0067$  \\
$I$         & $0.1065\pm0.0380$     & $-1.0456\pm0.4285$    & $0.0713\pm0.0264$     & $0.0188\pm0.0170$  \\
$z$         & $0.5406\pm0.0539$     & $-0.8770\pm0.6547$    & $0.1153\pm0.0432$     & $0.0175\pm0.0184$  \\
$J$         & $-0.1490\pm0.0153$    & $-1.7138\pm0.1834$    & $0.0385\pm0.0081$     & $0.0058\pm0.0017$  \\
$H$         & $-0.3509\pm0.0148$    & $-2.1936\pm0.1752$    & $0.0312\pm0.0068$     & $0.0060\pm0.0015$  \\
$K$         & $-0.3472\pm0.0160$    & $-2.4599\pm0.1849$    & $0.0498\pm0.0089$     & $0.0071\pm0.0019$  \\
$W1$        & $-0.4703\pm0.0112$    & $-2.1968\pm0.1252$    & $0.0032\pm0.0020$     & $0.0050\pm0.0013$  \\
$W2$        & $-0.4583\pm0.0112$    & $-2.2337\pm0.1249$    & $0.0055\pm0.0018$     & $0.0053\pm0.0016$  \\
$W3$        & $-0.4924\pm0.0119$    & $-2.3026\pm0.1342$    & $0.0227\pm0.0036$     & $0.0350\pm0.0291$ \\
\hline
\label{tab:plr_parameters}
\end{tabular}
\end{minipage}
\end{table*}

\subsection[]{MCMC fitting details}\label{sec:mcmc}

\begin{figure*}
	\centering
	\includegraphics{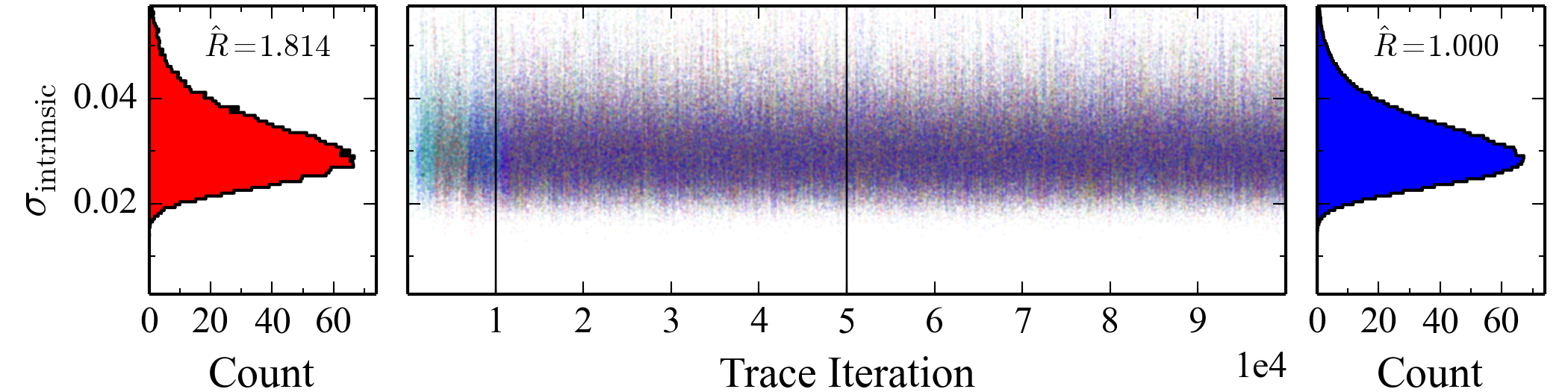}
	\caption{MCMC traces of intrinsic scatter, $\sigma_{\rm intrinsic}$, for the $H$ waveband. All seven traces of 100,000 samples each are plotted simultaneously, coloured by trace. The left panel shows the normalized histogram of the first 10,000 samples from each trace and the right panel shows the normalised histogram of the last 50,000 samples from each trace. In each histogram panel the Gelman-Rubin convergence diagnostic, $\hat{R}$, (\protect\citealt{gelman_rubin_convergence}) is given. $\hat{R}$ should converge to 1 and traces are generally considered converged when $\hat{R} \lesssim 1.1$. The first 50,000 samples are rejected as burn-in and the last 50,000 samples are considered to be drawn from the converged posterior distribution.}
	\label{fig:sigma_trace}
\end{figure*}

\begin{figure*}
	\centering
	\includegraphics{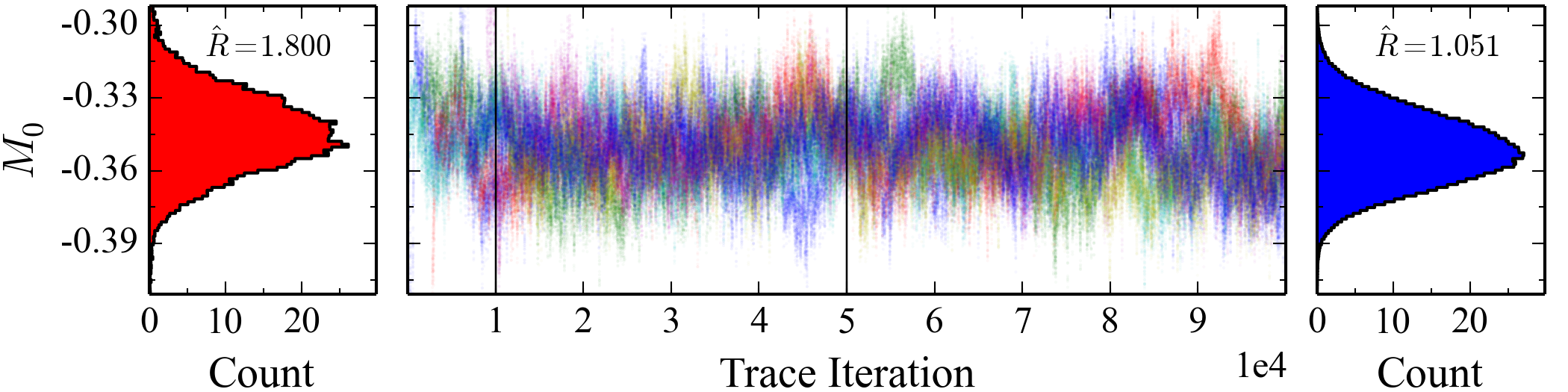}
	\caption{MCMC traces of $M_0$ for the $H$ waveband. Panels formatted as in Fig.\,\ref{fig:sigma_trace}.}
	\label{fig:M_0_trace}
\end{figure*}

\begin{figure*}
	\centering
	\includegraphics{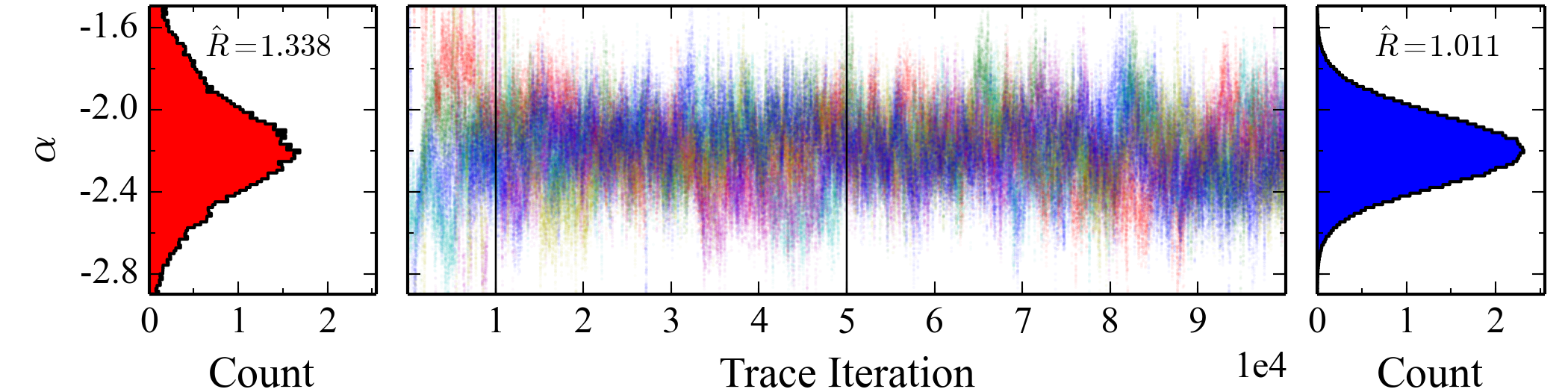}
	\caption{MCMC traces of $\alpha$ for the $H$ waveband. Panels formatted as in Fig.\,\ref{fig:sigma_trace}.}
	\label{fig:alpha_trace}
\end{figure*}

\begin{figure*}
	\centering
	\includegraphics{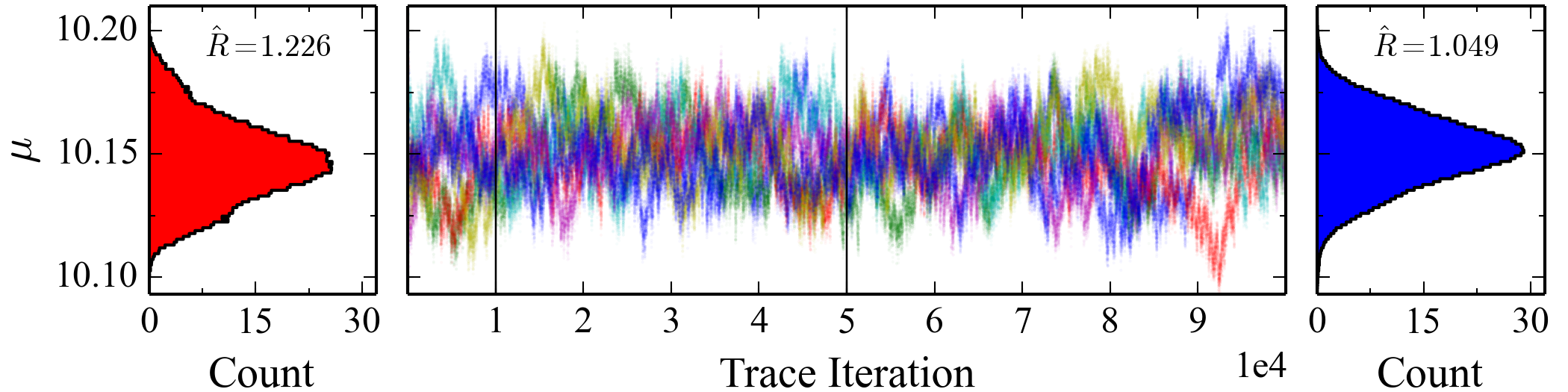}
	\caption{MCMC traces of $\mu$ for ABUMa. Panels formatted as in Fig.\,\ref{fig:sigma_trace}.}
	\label{fig:mu_trace}
\end{figure*}

\begin{figure*}
	\centering
	\includegraphics{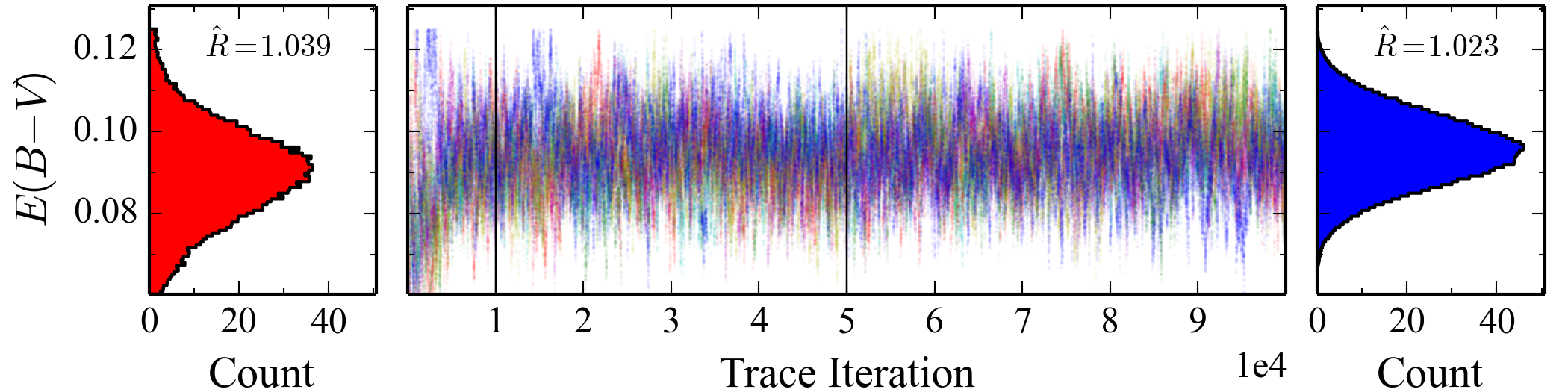}
	\caption{MCMC traces of  ${E(B-V)}$ for ABUMa. Panels formatted as in Fig.\,\ref{fig:sigma_trace}.}
	\label{fig:ebv_trace}
\end{figure*}

Seven MCMC sampling traces of the model fit were produced, each iterating 25,200,000 steps and thinned by a factor of 252 to result in traces with 100,000 iterations. As an illustrative example, Figs.\,\ref{fig:sigma_trace}, \ref{fig:M_0_trace}, and \ref{fig:alpha_trace} show trace plots for the $H$-band $\sigma_{\rm intrinsic}$,  $M_0$, and $\alpha$, respectively. Additionally, trace plots for the $\mu$ and  ${E(B-V)}$ of ABUMa are shown respectively in Figs.\,\ref{fig:mu_trace} and \ref{fig:ebv_trace}.

The traces are considered to be converged after 50,000 iterations, and these converged portions of each of the seven traces are combined to form a posterior distribution for each model parameter of 350,000 samples. Convergence is verified by computing the Gelman-Rubin multiple sequence convergence diagnostic, $\hat{R}$, (\protect\citealt{gelman_rubin_convergence}) and ensuring $\hat{R} \lesssim 1.1$ in the portion of the chains considered to be converged. The Gelman-Rubin diagnostic factor is the square root of the weighted sum of the within chain variance, $W$, and between chain variance, $B$, divided by the within chain variance. Here,
\begin{equation}
\label{eqn:GR_diagnostic}
\hat{R} = \sqrt{\frac{\left(1-\frac{1}{n}\right)W + \left(\frac{1}{n}\right)B}{W}},
\end{equation}
where $n$ is the of length each chain. In Figs.\,\ref{fig:sigma_trace} through \ref{fig:ebv_trace} the Gelman-Rubin diagnostic is displayed for the first 10,000 iterations (demonstrating the lack of convergence early in the MCMC sampling chain) and also for the final 50,000 iterations (where the traces are considered to be converged).

\subsection[]{Zero point and slope joint distributions}

\begin{figure}
	\centering
	\includegraphics{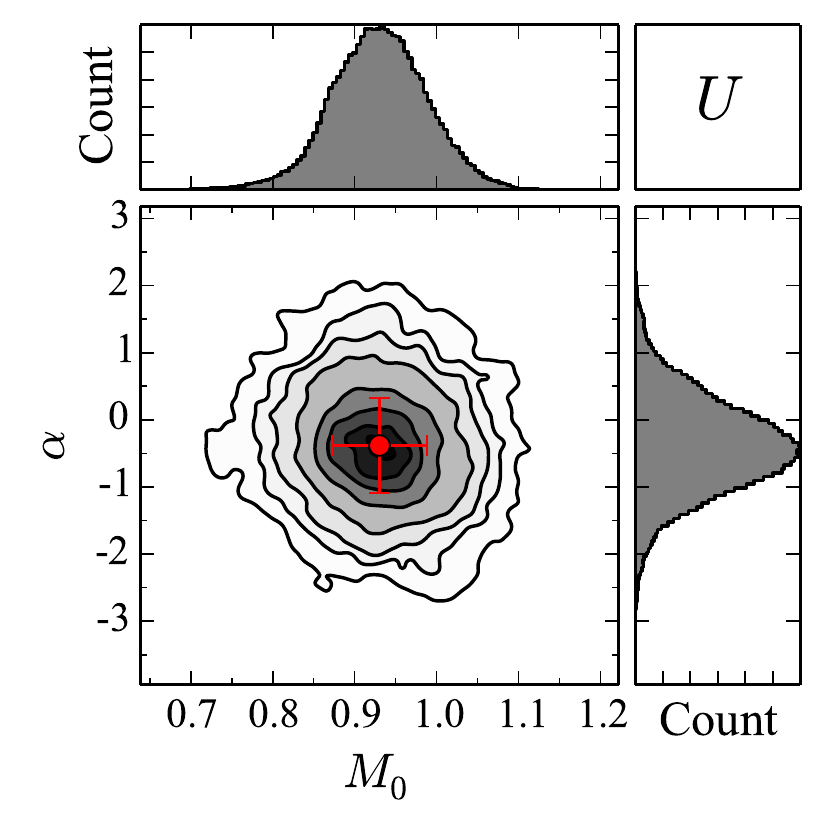}
	\caption{Contour density plot and histograms for the $U$ band period--magnitude relation magnitude intercept ($M_0$) and slope ($\alpha$). The red circle with associated error bars shows the means and standard deviations of the posterior $M_0$ and $\alpha$ distributions.}
	\label{fig:U_contour}
\end{figure}

\begin{figure}
	\centering
	\includegraphics{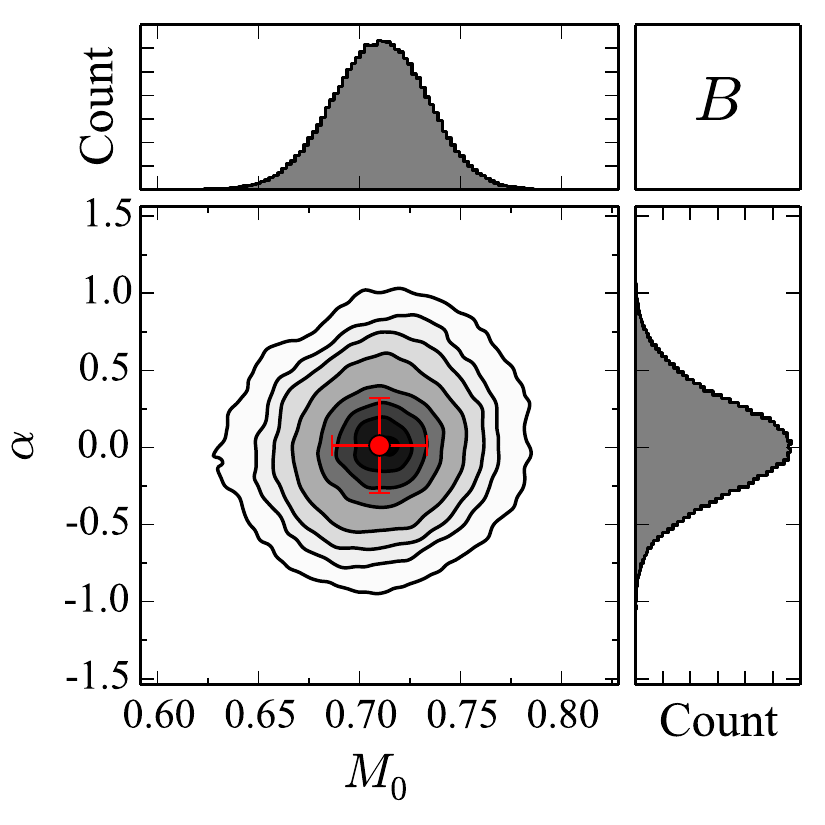}
	\caption{Contour density plot and histograms for the $B$ band period--magnitude relation magnitude intercept ($M_0$) and slope ($\alpha$). The red circle with associated error bars shows the means and standard deviations of the posterior $M_0$ and $\alpha$ distributions.}
	\label{fig:B_contour}
\end{figure}

\begin{figure}
	\centering
	\includegraphics{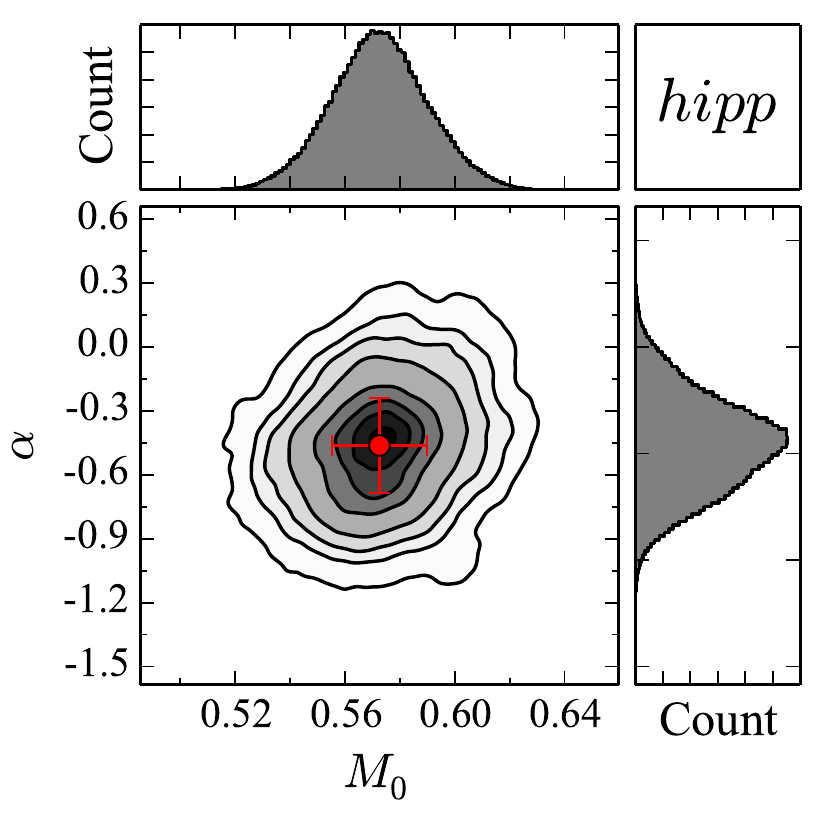}
	\caption{Contour density plot and histograms for the $hipp$ band period--magnitude relation magnitude intercept ($M_0$) and slope ($\alpha$). The red circle with associated error bars shows the means and standard deviations of the posterior $M_0$ and $\alpha$ distributions.}
	\label{fig:hipp_contour}
\end{figure}

\begin{figure}
	\centering
	\includegraphics{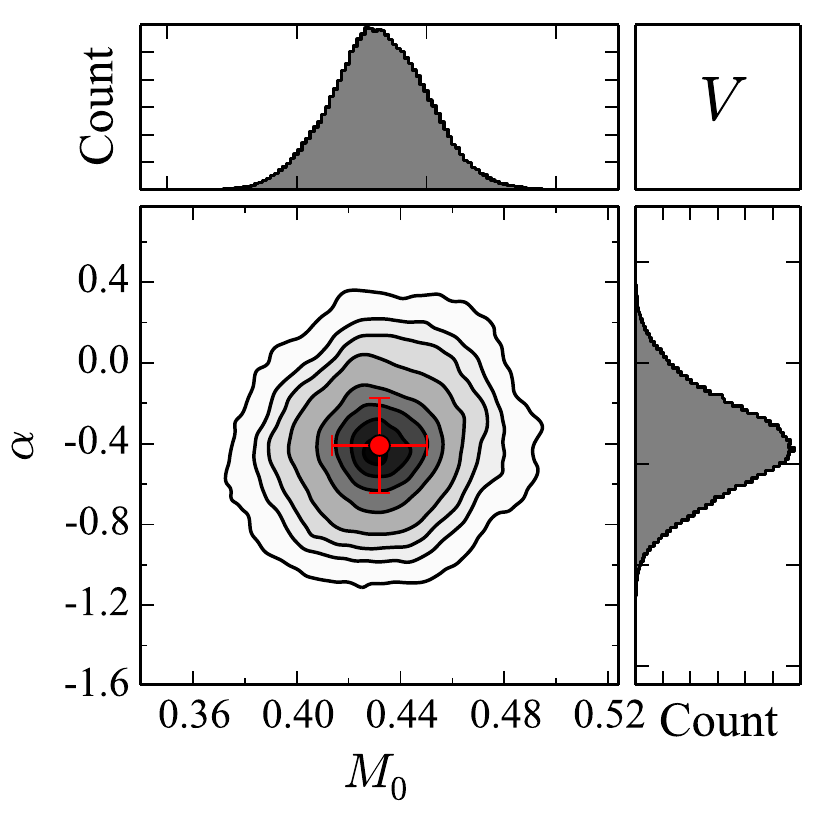}
	\caption{Contour density plot and histograms for the $V$ band period--magnitude relation magnitude intercept ($M_0$) and slope ($\alpha$). The red circle with associated error bars shows the means and standard deviations of the posterior $M_0$ and $\alpha$ distributions.}
	\label{fig:V_contour}
\end{figure}

\begin{figure}
	\centering
	\includegraphics{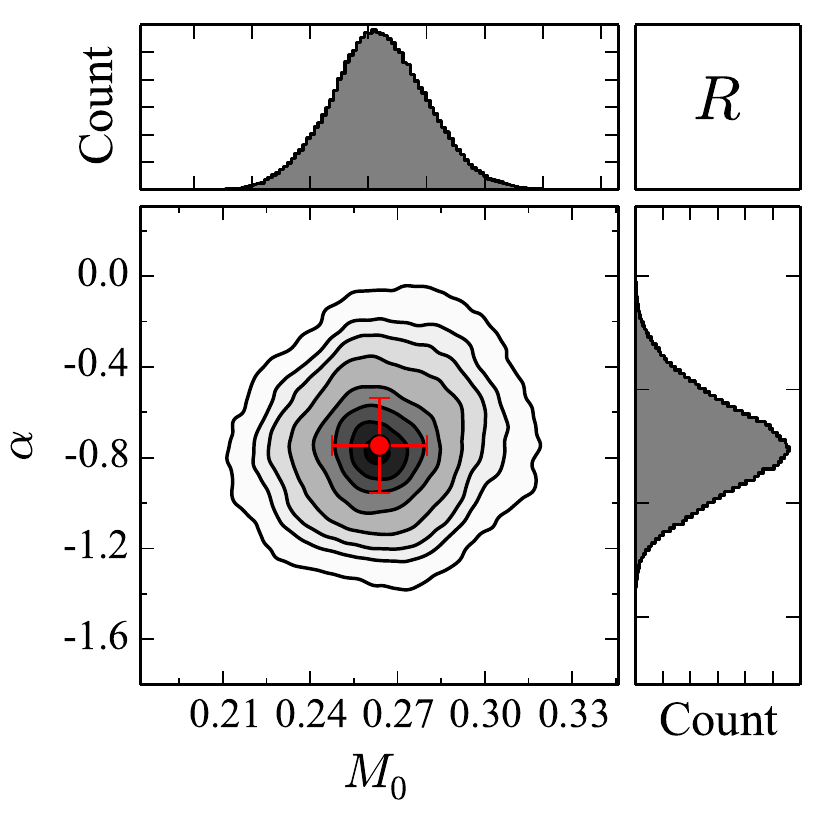}
	\caption{Contour density plot and histograms for the $R$ band period--magnitude relation magnitude intercept ($M_0$) and slope ($\alpha$). The red circle with associated error bars shows the means and standard deviations of the posterior $M_0$ and $\alpha$ distributions.}
	\label{fig:R_contour}
\end{figure}

\begin{figure}
	\centering
	\includegraphics{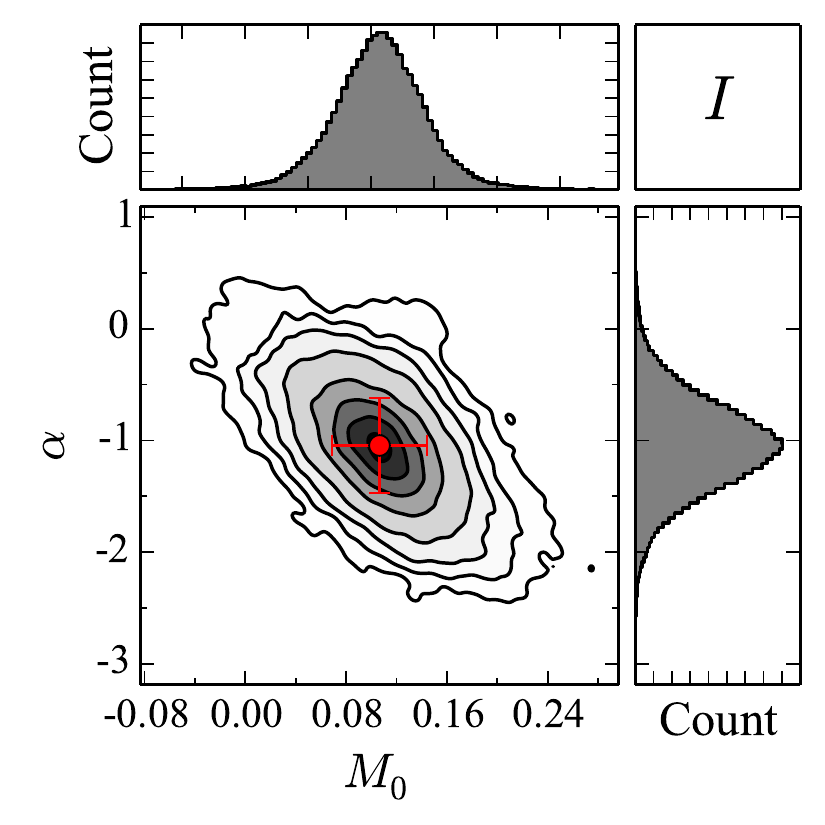}
	\caption{Contour density plot and histograms for the $I$ band period--magnitude relation magnitude intercept ($M_0$) and slope ($\alpha$). The red circle with associated error bars shows the means and standard deviations of the posterior $M_0$ and $\alpha$ distributions. The exhibited correlation in these parameters for $I$ band is caused by the lopsided period distribution of the RR Lyrae variables for which $I$ band data was obtained (c.f. Fig.\,\ref{fig:plrs}, only three of nine stars have $P<P_0$).}
	\label{fig:I_contour}
\end{figure}

\begin{figure}
	\centering
	\includegraphics{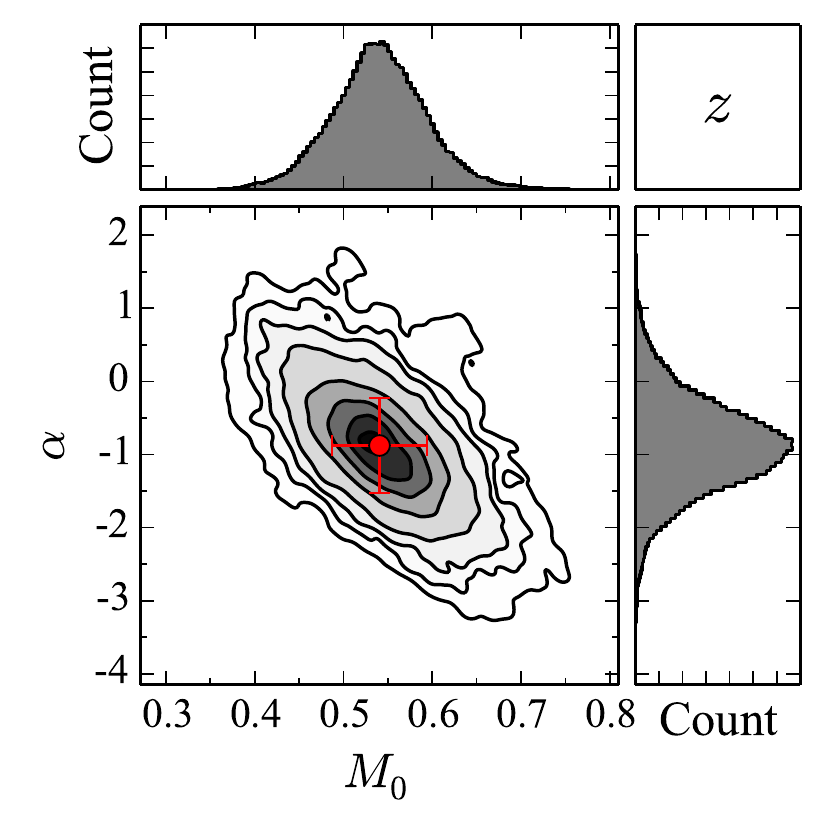}
	\caption{Contour density plot and histograms for the $z$ band period--magnitude relation magnitude intercept ($M_0$) and slope ($\alpha$). The red circle with associated error bars shows the means and standard deviations of the posterior $M_0$ and $\alpha$ distributions. As in Fig.\,\ref{fig:I_contour}, a correlation in these parameters is obvious. The explanation, an uneven period distribution about $P_0$ in the subset of RR Lyrae stars for which $z$ band data was obtained, is the same as for $I$ band (Fig.\,\ref{fig:I_contour}).}
	\label{fig:z_contour}
\end{figure}

\begin{figure}
	\centering
	\includegraphics{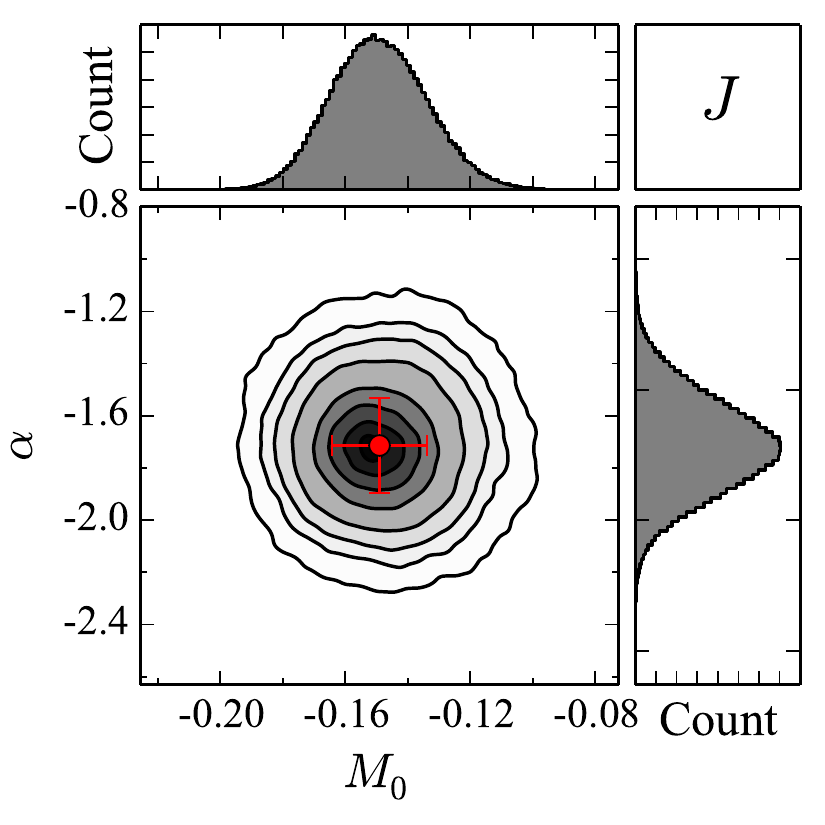}
	\caption{Contour density plot and histograms for the $J$ band period--magnitude relation magnitude intercept ($M_0$) and slope ($\alpha$). The red circle with associated error bars shows the means and standard deviations of the posterior $M_0$ and $\alpha$ distributions.}
	\label{fig:J_contour}
\end{figure}

\begin{figure}
	\centering
	\includegraphics{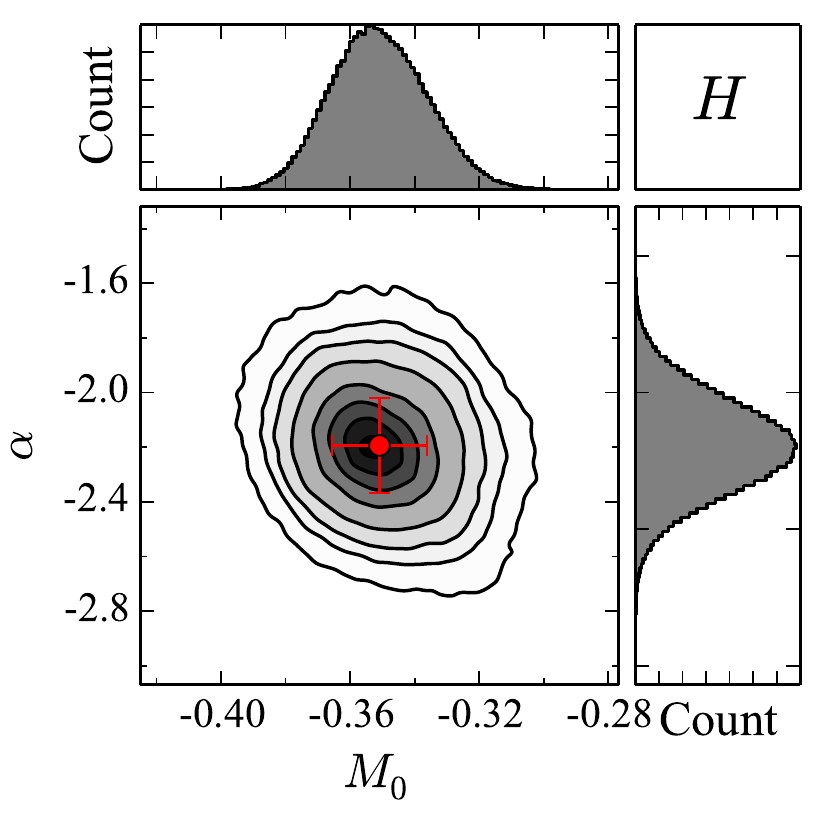}
	\caption{Contour density plot and histograms for the $H$ band period--magnitude relation magnitude intercept ($M_0$) and slope ($\alpha$). The red circle with associated error bars shows the means and standard deviations of the posterior $M_0$ and $\alpha$ distributions.}
	\label{fig:H_contour}
\end{figure}

\begin{figure}
	\centering
	\includegraphics{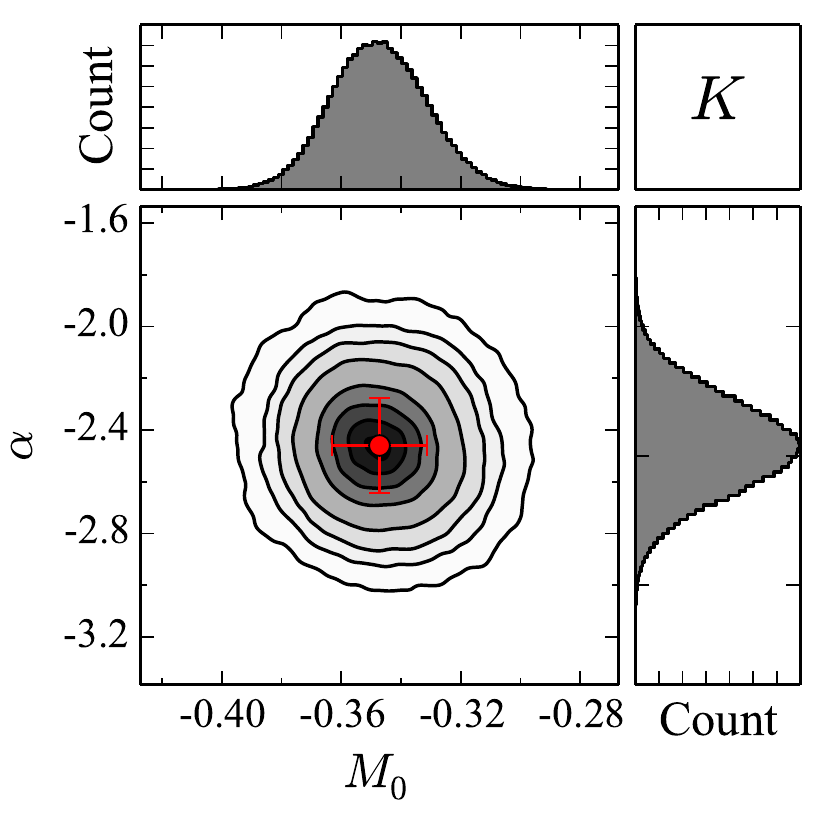}
	\caption{Contour density plot and histograms for the $K$ band period--magnitude relation magnitude intercept ($M_0$) and slope ($\alpha$). The red circle with associated error bars shows the means and standard deviations of the posterior $M_0$ and $\alpha$ distributions.}
	\label{fig:K_contour}
\end{figure}

\begin{figure}
	\centering
	\includegraphics{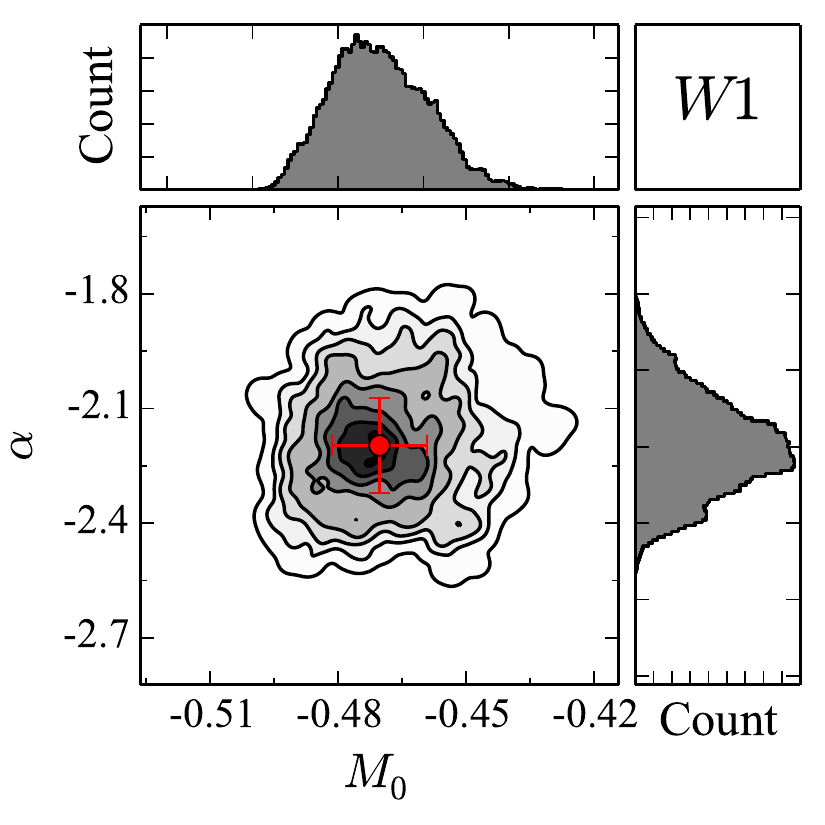}
	\caption{Contour density plot and histograms for the $W1$ band period--magnitude relation magnitude intercept ($M_0$) and slope ($\alpha$). The red circle with associated error bars shows the means and standard deviations of the posterior $M_0$ and $\alpha$ distributions.}
	\label{fig:W1_contour}
\end{figure}

\begin{figure}
	\centering
	\includegraphics{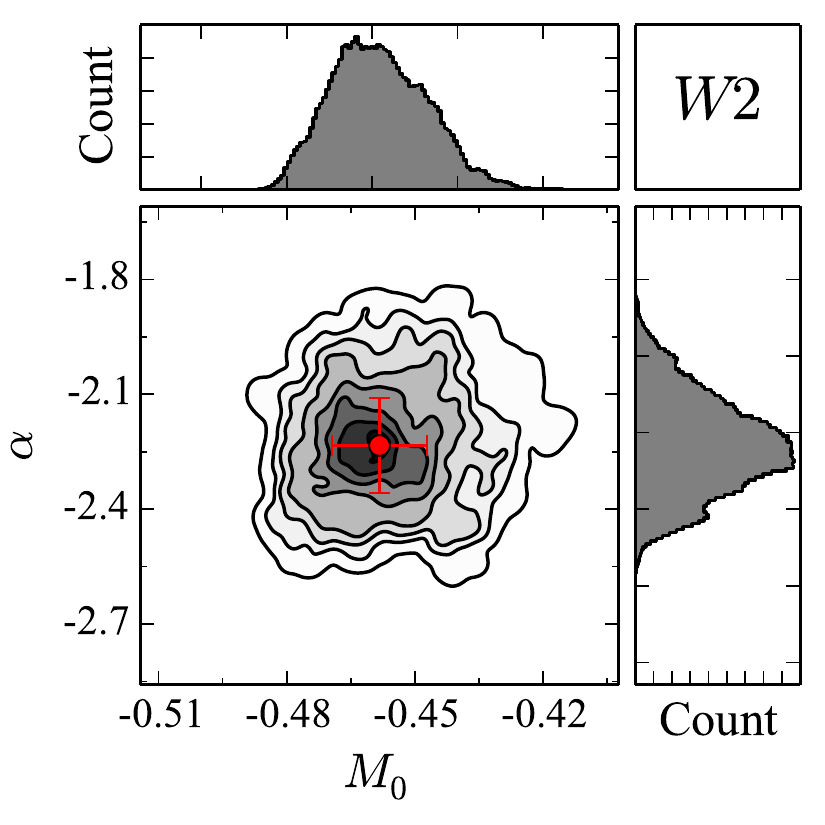}
	\caption{Contour density plot and histograms for the $W2$ band period--magnitude relation magnitude intercept ($M_0$) and slope ($\alpha$). The red circle with associated error bars shows the means and standard deviations of the posterior $M_0$ and $\alpha$ distributions.}
	\label{fig:W2_contour}
\end{figure}

\begin{figure}
	\centering
	\includegraphics{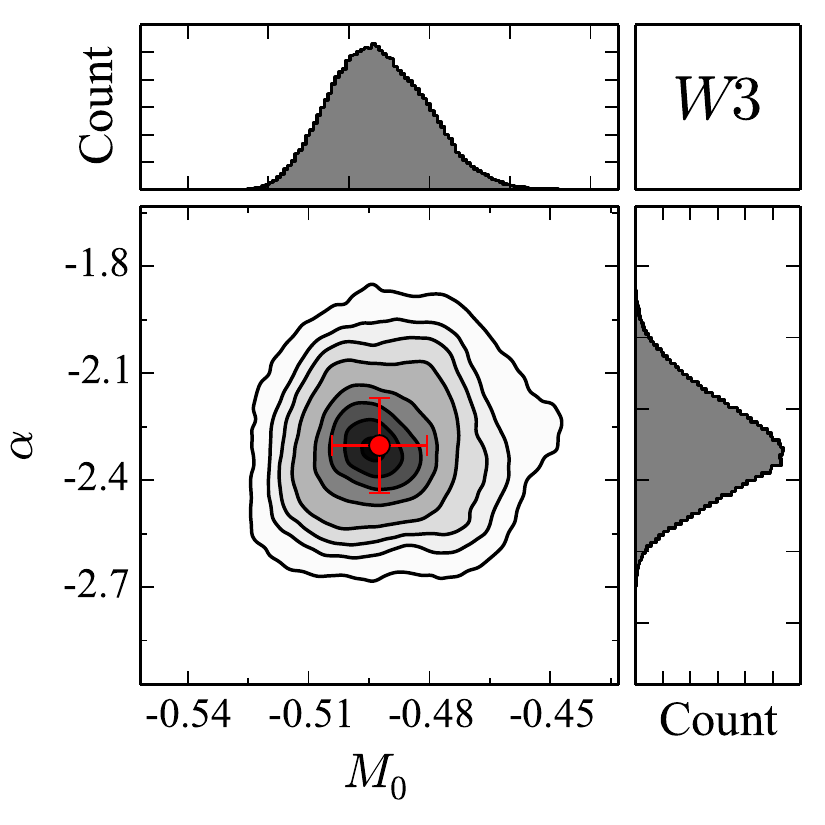}
	\caption{Contour density plot and histograms for the $W3$ band period--magnitude relation magnitude intercept ($M_0$) and slope ($\alpha$). The red circle with associated error bars shows the means and standard deviations of the posterior $M_0$ and $\alpha$ distributions.}
	\label{fig:W3_contour}
\end{figure}

One significant advantage of a Bayesian approach to linear regression over frequentist methods is that the posterior model parameters are sampled from final joint distributions. Thus, any covariance in the distributions is accurately recorded and the traditional assumption of Gaussian behaviour is not necessary, but can instead be tested. Indeed, the posterior $M_0$ and $\alpha$ distributions are generally well-approximated by Gaussians, but some waveband-specific pairs exhibit covariance. Figs.\,\ref{fig:U_contour} through \ref{fig:W3_contour} display the posterior contour density plots and histograms for the zero point and slope of the 13 waveband-specific period--magnitude relations.

The pronounced covariance between $M_0$ and $\alpha$ observed for the $I$ and $z$ wavebands is primarily caused by the lop-sided distribution of the periods of the RR Lyrae stars for which $I$- and $z$-band data were obtained. Only three of the nine stars observed in these wavebands have $P<P_0$, and thus the covariance between the linear regression intercept and slope was not well-removed.

\clearpage

\clearpage

\subsection[]{Period--magnitude relation plot}\label{PM_relation_plot}

\begin{figure*}
	\centering
	\includegraphics{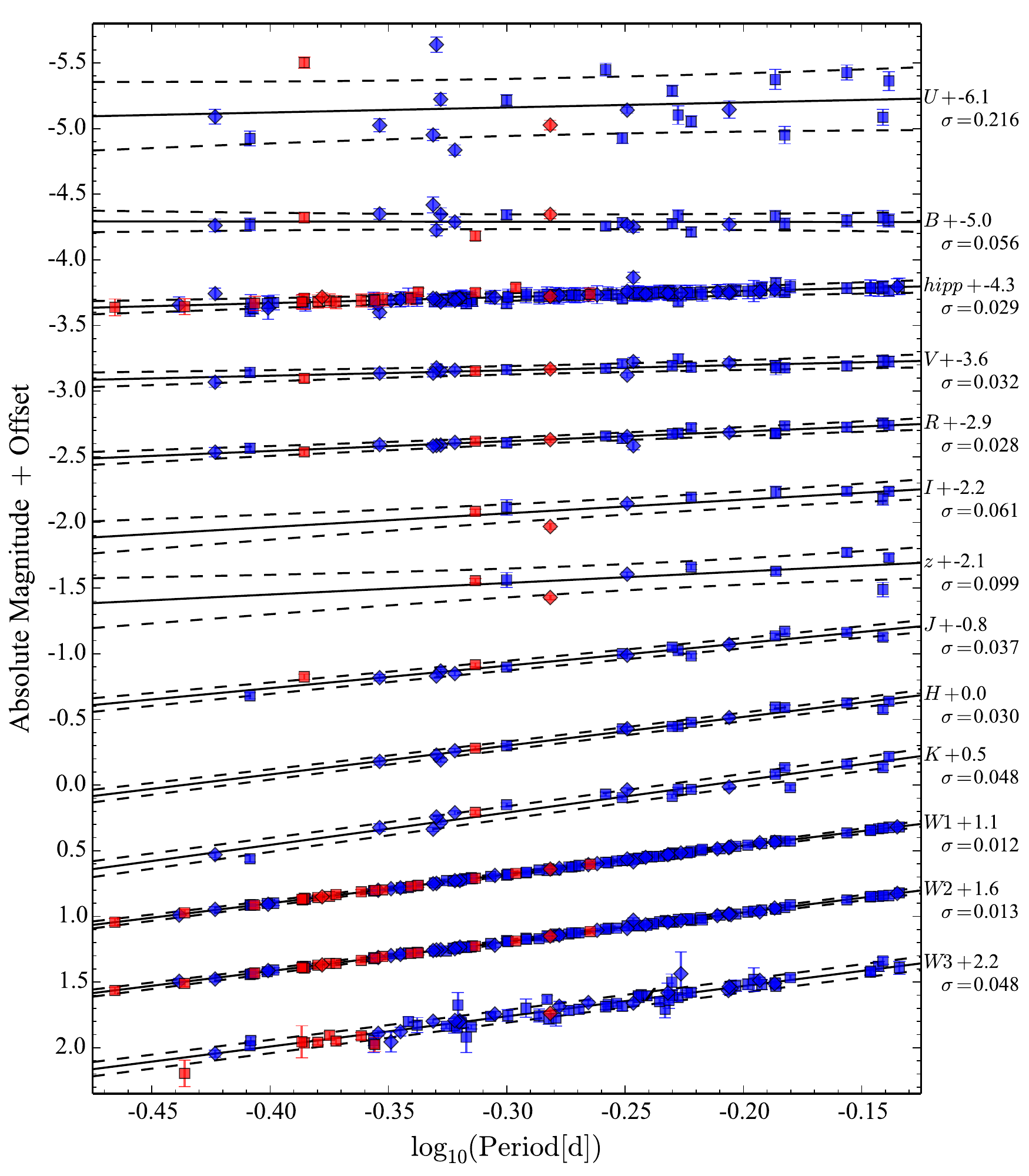}
	\caption{Multi-band period--magnitude relations. RRab stars are in blue, RRc stars in red. Blazhko-affected stars are denoted with diamonds, stars not known to exhibit the Blazhko effect are denoted with squares. Solid black lines are the best-fitting period--magnitude relations in each waveband and dashed lines indicate the 1-$\sigma$ prediction uncertainty for application of the best-fitting period--magnitude relation to a new star with known period. The noted scatter, $\sigma$ associated with each band in the figure, is the minimum prediction uncertainty, which is where the dashed line ``bowtie pinch'' around $P_0=0.52854$ d.}
	\label{fig:plrs}
\end{figure*}

Fig.\,\ref{fig:plrs} depicts the 13 period--magnitude relations in one large plot. The zero points of the relations are shifted vertically in the plot, as noted with the offsets given on the right hand side, to separate out the relations as displayed graphically. The solid black lines denote the best-fitting period--magnitude relations, and the dashed lines indicate the $1\sigma$ prediction uncertainty for application of the best-fitting period--magnitude relation to a new star with known period. Also noted on the right hand side is the minimum prediction uncertainty, here given simply as $\sigma$, which provides a sense for how accurately a single new RR Lyrae star's absolute magnitude can be predicted from a given band-specific period--magnitude relation. This value is the minimum vertical distance between the solid and dashed lines for each relation (which usually occurs around $P_0$). RRab stars are denoted with blue markers and RRc stars are shows in red. The plot contains 637 markers, one for each RR Lyrae light curve. 

In addition to illustrating the 13 period--magnitude relations, this plot can also be interpreted to show the waveband distribution of the 637 light curves in the calibration dataset. For example, it is evident that the space-based {\it Hipparcos} and {\it WISE} missions provide light curve data in their wavebands for most of the calibration sample, and also that only nine stars were observed in $I$ and $z$ (and that those nine are biased towards longer periods). The plot also provides a graphical display of the proportion of RRab versus RRc stars (blue versus red markers) in the sample, broken down by waveband.

A large fraction of RR Lyrae stars (at least 20 per cent, and likely significantly more) are affected by an amplitude modulation called the Blazhko effect. This effect manifests as a slow cyclic evolution of the light curve shape, with a period ranging from weeks to months (\citealt{1995CAS....27.....S}, chapter 5.2). The nature of the Blazhko effect, a second-order amplitude modulation, does not result in a significant impact on a star's mean-flux magnitude. In Fig.\,\ref{fig:plrs} Blazhko-affected stars, as identified via \url{http://www.univie.ac.at/tops/blazhko/Blazhkolist.html}, are shown with diamonds and stars without confirmed evidence of the Blazhko effect are shown with squares.  

Because of the longer-period nature of the effect, observational investigations of the RR Lyrae Blazhko effect require considerable telescope resources. To our knowledge, no such investigations have been carried out in near- or mid-infrared wavebands. That the amplitude distribution of RR Lyrae stars is significantly reduced in the near- and mid-infrared, as compared to optical bands, suggests that the magnitude of the Blazkho effect will also be diminished in the infrared \citep{2013arXiv1312.4643G}. However, observational studies are required to test this hypothesis. In the present analysis, and as indicated in Fig.\,\ref{fig:plrs}, there is no significant impact on the period--magnitude relation by the inclusion of Blazhko-affected stars in the fit. 

\begin{figure}
	\centering
	\includegraphics{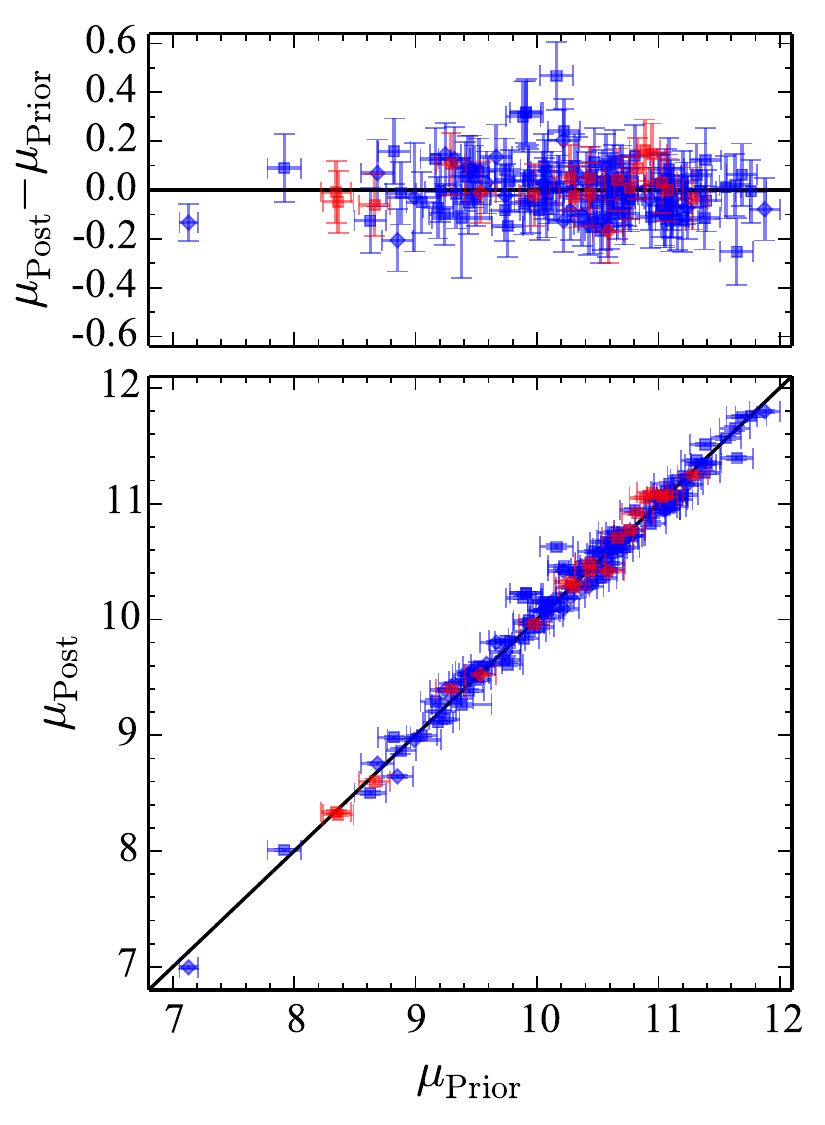}
	\caption{Prior versus posterior distance moduli, with residual plot in upper panel. Colours and symbols as in Fig.\,\ref{fig:plrs}. 112 out of the 134 (84 per cent) of the residual data points are within one error bar length of zero, indicating that the errors are slightly overestimated.}
	\label{fig:mu-mu}
\end{figure}

As a commonsense check on the period--magnitude relations of Fig.\,\ref{fig:plrs} and the applied  simultaneous Bayesian linear regression fitting method, a plot of the prior distance moduli versus the posterior distance moduli for the calibration sample is provided in Fig.\,\ref{fig:mu-mu}. Any bias or a strongly non-normal distribution of the $\mu_{\rm Post} - \mu_{\rm Prior}$ residuals would indicate overfitting. Since the distance modulus is treated as a model parameter to be fit, it is very important that the fitting method respects the original prior distance modulus values.
Fig.\,\ref{fig:mu-mu} shows that the prior and posterior distances are in very good agreement. Specifically, 112 out of the 134 calibrators (84 per cent) have $\mu_{\rm Post} - \mu_{\rm Prior}$ residuals that lie within one-errorbar length of zero.
The errors on the posterior distance moduli may thus be slightly overestimated. The null hypothesis that the posterior-prior distance modulus residuals are drawn from a standard normal distribution is accepted by a  Kolmogorov-Smirnov test with $p=0.2$.

In Fig.\,\ref{fig:mu-mu} RRLyr itself is the star with the lowest distance modulus. The fitted posterior distance modulus of RRLyr is ${\mu_{\rm Post} = 6.9962 \pm 0.0143}$ with a prior distance modulus, derived from the measured HST parallax, of ${\mu_{\rm Prior} = 7.130 \pm 0.075}$. For RRLyr, the residual significance is $-1.75\sigma$.

\section[]{Further Discussion of the Fits}\label{discussion}

The complex model used in the period--magnitude relation fits (described above in Section\,\ref{derivations}), which newly incorporates colour excess and intrinsic scatter, allows for a deeper analysis of the results. In the following subsections we present the fit results as spectral energy distributions (SEDs), compare the fitted period--magnitude relation intrinsic scatter and mean photometric error as a function of wavelength, analyse the colour excess results more closely, and discuss the evolution of period--magnitude relation slope with wavelength. 

\subsection[]{RR Lyrae spectral energy distributions}

\begin{figure}
	\centering
	\includegraphics{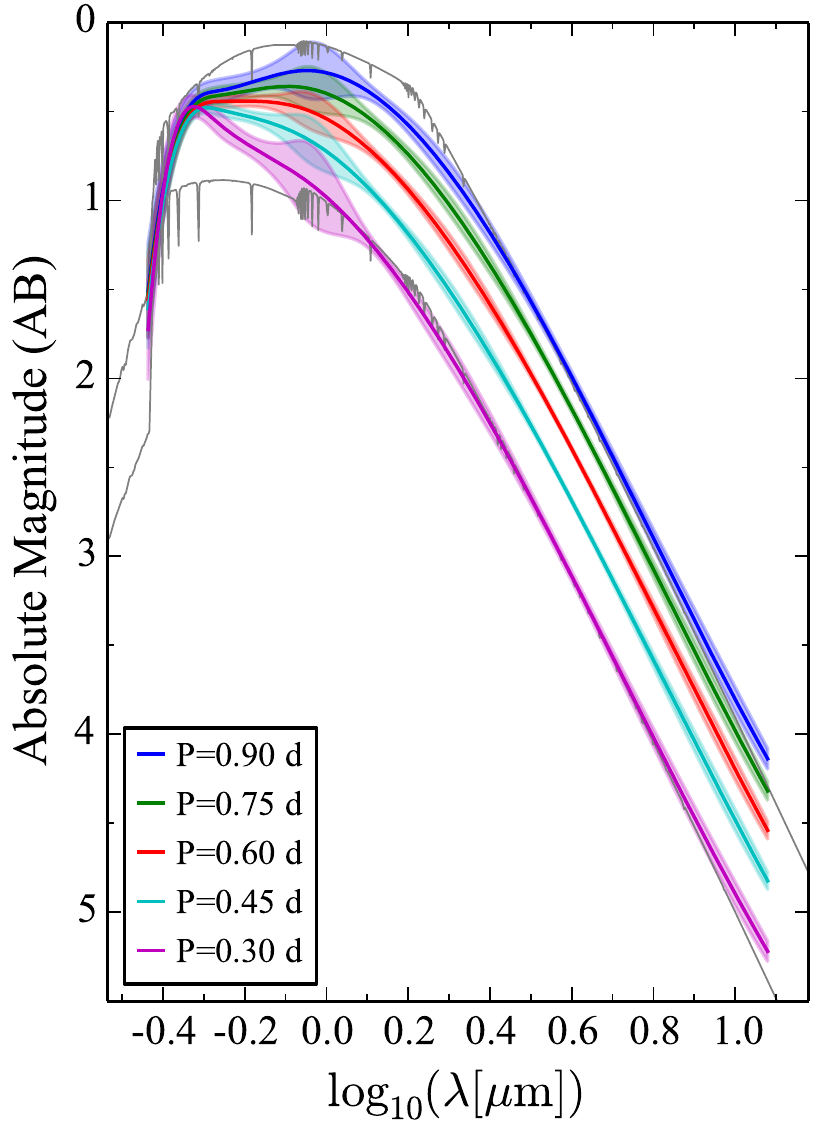}
	\caption{Spectral energy distributions of RR Lyrae stars with assorted fundamental oscillation periods, derived from the period--magnitude relations of Fig.\,\ref{fig:plrs}. The individual data points are omitted to permit clearer perception of the spline interpolation curves. Note the large width around 1 $\mu$m, which is caused by the less tightly constrained $I$- and $z$-band period--magnitude relations. The light gray lines are model stellar spectra from Castelli and Kurucz Atlas 2004 accessed from Space Telescope Science Institute FTP (\url{http://www.stsci.edu/ftp/cdbs/grid/ck04models}). The models are meant to bracket the RR Lyrae parameter space in radius and effective temperature. The brighter model spectrum corresponds to $T_{\rm eff}=6250$ K and $R=7~{\rm R}_\odot$, and the dimmer spectrum to $T_{\rm eff}=7000$ K and $R=4~{\rm R}_\odot$. Both models are for $\log g = 2.5$ and $\log Z = -2.5$. This temperature and radius range is consistent with previous work: c.f.\,\protect\cite{1995CAS....27.....S}.}
	\label{fig:seds}
\end{figure}

The period--magnitude relation fits provide the absolute magnitudes (at time of mean-magnitude) of the typical RR Lyrae star in 13 wavebands as a function of period. Another way to present, and think about, this result is by converting the fits to SEDs for RR Lyrae stars at selected periods. This is demonstrated in Fig.\,\ref{fig:seds}, along with two model stellar spectra (light grey lines) selected with temperatures and radii to bracket the ranges of these parameters inferred in the RR Lyrae population.

This plot of SEDs for RR Lyrae stars of various fundamental periods illustrates why the period--magnitude relations at optical wavebands (near the SED peak) have a shallower slope than at infrared wavebands (along the Rayleigh-Jeans tail). The vertical distance between two SEDs tracks with the slope of the period--magnitude relation. This vertical distance between the brightest SED (longest period RR Lyrae star) and the dimmest SED (shortest period) is effectively zero shortward $R$-band, and then this distance increases with increasing wavelength until the SEDs become nearly parallel in the Rayleigh-Jeans tail. This near-parallel property of the SEDs in the infrared graphically explains why the period--magnitude relation slope approaches an asymptote with increasing wavelength. 

\subsection[]{Intrinsic scatter and photometric error}

\begin{figure}
	\centering
	\includegraphics{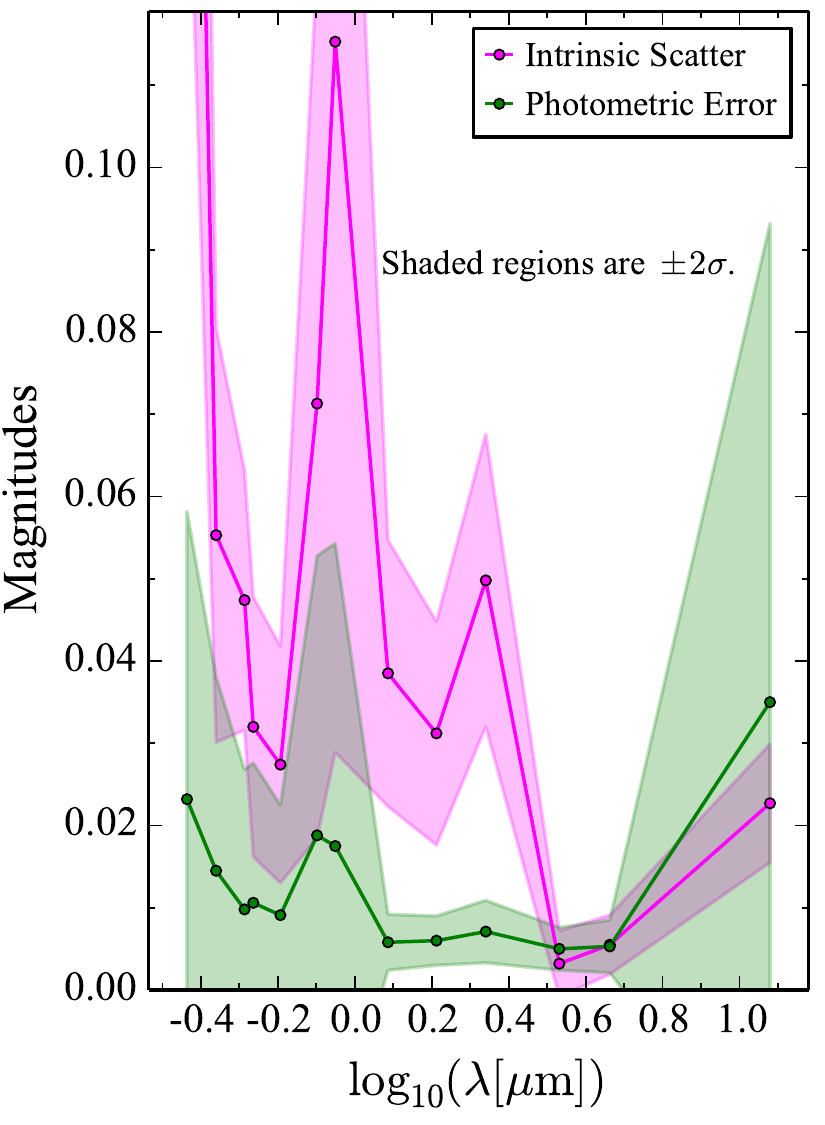}
	\caption{Photometric uncertainty and fitted intrinsic scatter as a function of wavelength. Note that the intrinsic scatter at the mid-infrared wavebands ($W1$, $W2$, and $W3$) is of similar value to the average photometric errors, whereas at shorter wavelengths the intrinsic scatter dominates and is the primary source of distance prediction uncertainty in those waveband-specific period--magnitude relations.}
	\label{fig:sigmas}
\end{figure}

As described in Section\,\ref{derivations} the model used in the period--magnitude relation fits allows for investigation of the intrinsic scatter, and of particular importance is the comparison between intrinsic scatter and the photometric error on the mean-flux magnitude measurements. Fig.\,\ref{fig:sigmas} shows both intrinsic scatter and mean photometric error as a function of wavelength. At any given wavelength, the maximum of the intrinsic scatter and mean photometric error provides a floor to how tightly the resultant period--magnitude relation can be constrained (and, in effect, sets the precision limit of distance measurements).

If the photometric error dominates over the intrinsic scatter, then a tighter period--magnitude relation can be derived by collecting better light curve data (i.e., with more sensitive instruments and/or more observation epochs). However, if the intrinsic scatter exceeds the mean photometric error already achieved, then the path towards a tighter period--magnitude relation is not as direct. In this latter case, the period--magnitude relation scatter can be reduced slowly via the augmentation of the calibration sample, but the intrinsic scatter will always dictate the minimum absolute magnitude uncertainty when applying the relation to new stars. The inclusion of a spectroscopically derived metallicity as an additional model parameter could, of course, serve to reduce the intrinsic scatter \citepeg{2006ARA&A..44...93S} but we expressly have used a model based upon photometry alone.

Fig.\,\ref{fig:sigmas} shows that the intrinsic scatter exceeds the photometric error for all wavebands except $W1$ and $W2$. This explains why the minimum prediction uncertainty given in Fig.\,\ref{fig:plrs}, $\sigma$ along the right hand side, is lowest for these wavebands. Furthermore, this finding indicates that continued development and application of RR Lyrae period--magnitude relations at wavebands between 3 and 5 $\mu$m will produce the tightest absolute magnitude constraints.

\subsection[]{Colour excess results}

A major improvement to the model fit in the present multi-band period--magnitude relations derivation is simultaneously fitting for colour excess to each of the calibration stars. This is not feasible with light curve data for only one wavelength regime (such as the work published in \citealt{2011ApJ...738..185K} and \citealt{2014MNRAS.440L..96K}). However, the present investigation spans the optical, near-infrared, and mid-infrared wavelength regimes, and this enables colour excess to be treated as a model parameter.

\begin{figure}
	\centering
	\includegraphics{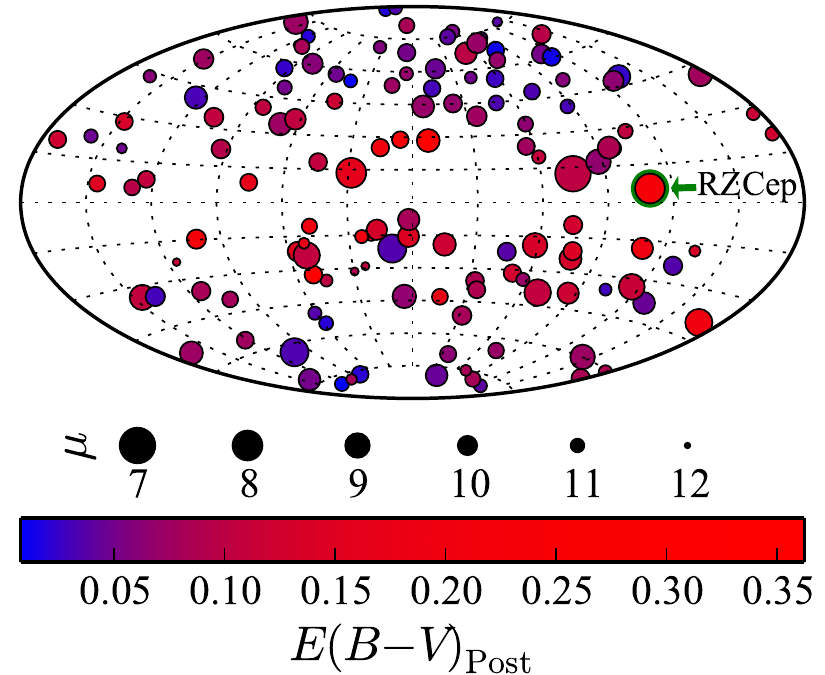}
	\caption{Aitoff projection in galactic coordinates of the RR Lyrae stars with symbols coloured by posterior fitted colour excess, $E(B-V)_{\rm Post}$, and sized by posterior fitted distance modulus, $\mu$. The centre of the figure is directed towards the Galactic Centre. The distance and colour excess of RZCep, a star rejected from the calibration sample and used as an example application in Section\,\ref{application} of the period--magnitude relations, is represented by the marker outlined in green and indicated by the label and green arrow.}
	\label{fig:ebv_residual_aitoff}
\end{figure}

An all-sky visualisation in Galactic coordinates of the fitted colour excess values, as well as the distance modulus, is shown in Fig.\,\ref{fig:ebv_residual_aitoff}. The colourbar is purposefully asymmetric to better conform to the dynamic range of the colour excess values (most of the values are near 0.08\,mag, and very few fall between 0.2\,mag and 0.35\,mag). An obvious feature of this skymap is the lack of RR Lyrae stars near the Galactic plane. This was enforced by the sample selection criteria discussed in Section\,\ref{data_desc}. A second visual trend is that the stars closer to the plane generally have higher colour excess values than those nearer the poles due to higher concentrations of interstellar dust near the Galactic plane.

\begin{figure}
	\centering
	\includegraphics{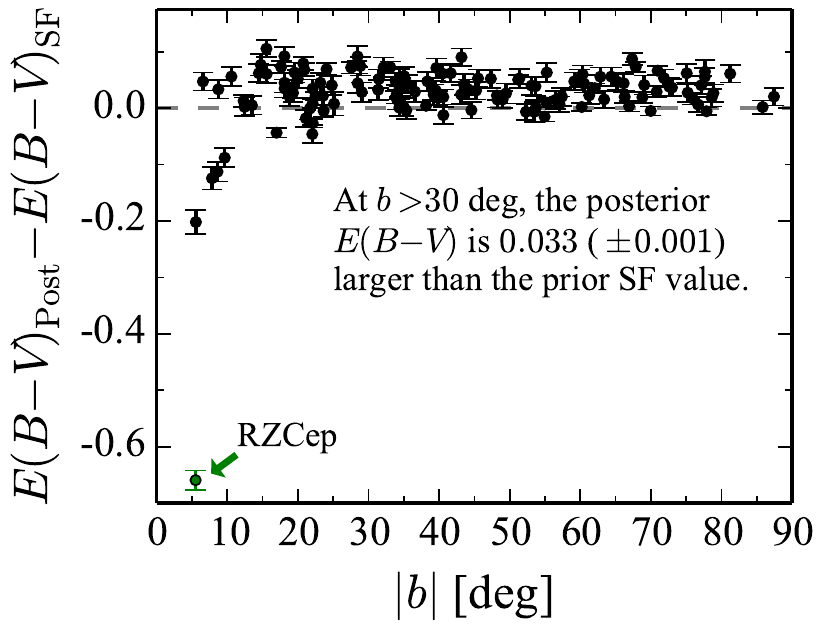}
	\caption{Change in colour excess,  ${E(B-V)}$, as a function of Galactic latitude, $b$. RZCep, a star rejected from the calibration sample and used as an example application in Section\,\ref{application} of the period--magnitude relations, is represented by the green marker. The large colour excess residual of RZCep indicates that most of the dust contributing to the ${E(B-V)_{\rm SF}=0.9054}$ value along its line of sight actually resides behind the star, as expected for a nearby ($d=405.4 \pm 2.3$ pc) star at a galactic latitude of $5.5^\circ$.}
	\label{fig:ebv_residual}
\end{figure}

To further explore the fitted colour excess values, Fig.\,\ref{fig:ebv_residual} shows the residual  ${E(B-V)}$ as a function of the absolute value of Galactic latitude, $b$. It is expected that the prior $E(B-V)_{\rm SF}$ values at low Galactic latitude are greater than the fitted posterior values, since the prior values represent the full colour excess expected along a line of sight to infinite distance whereas the posterior values follow a line of sight that terminates at the star (which presumably lies in front of much of the dust that contributes to the prior colour excess value). This expectation is indeed seen to hold in Fig.\,\ref{fig:ebv_residual} for galactic latitudes less than about 15 deg.

An unexpected feature of Fig.\,\ref{fig:ebv_residual} is that the mean residual colour excess does not settle around zero at high galactic latitude. At high latitude a calibrator RR Lyrae star should be behind most of the interstellar dust, and thus the posterior value should approach the prior $E(B-V)_{\rm SF}$ value. However, the mean residual at $b>30^\circ$ is $0.033 \pm 0.001$ (with scatter about the mean of 0.024), indicating that either the $E(B-V)_{\rm SF}$ values have a systematic bias, the calibrator RR Lyrae stars are more likely to lie behind more dust than nearby lines of sight as measured in the \cite{1998ApJ...500..525S} dust map, the value of $R_V=3.1$ is systematically incorrect, or some combination of all three.

\subsection[]{Period--magnitude relation slope}

\begin{figure}
	\centering
	\includegraphics{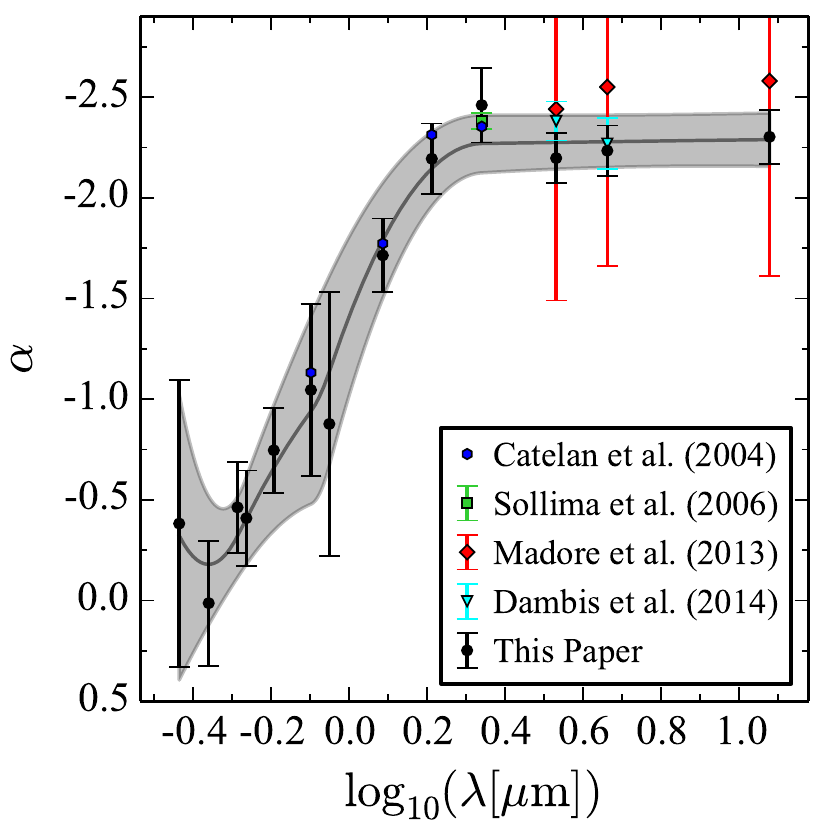}
	\caption{Period--magnitude relation slope ($\alpha$) as a function of wavelength. Gray band is a spline interpolation of the new calibrations presented in this paper.}
	\label{fig:slope}
\end{figure}

Fig.\,\ref{fig:slope} depicts the period--magnitude relation slope as a function of wavelength for the results from this work and other recent studies. In particular, \cite{2004ApJS..154..633C} produced theoretical calibrations of the period--magnitude relation at $I$, $J$, $H$, and $K$. \cite{2006MNRAS.372.1675S} provides a $K$-band relation derived from observations of globular clusters. \cite{2013ApJ...776..135M} and \cite{2014MNRAS.439.3765D} both derive mid-infrared relations using {\it WISE} data, the former using four calibrators with HST parallax measurements and the latter using RR Lyrae stars detected by {\it WISE} in globular clusters.

As indicated by the RR Lyrae SEDs (c.f. Fig.\,\ref{fig:seds}), the period--magnitude relation slope is observed to asymptote with increasing wavelength to a value around $-2.3$. Note that the slope ($\alpha$) axis is plotted with lower (steeper) values nearer to the top of the figure. The figure also shows a turnover in the slope in the $B$-band, but the uncertainty in the $U$-band slope is too large to rule out the monotonic trend of increasing slope (decreasing in absolute value) with decreasing wavelength. There is a qualitative concordance of our results with other observational and theoretical work.

\section[]{Example Application}\label{application}

Light curve data were obtained for RZCep (RRc star with period 0.308645 d, or fundamentalised period 0.413484 d) in the $U$, $B$, $hipp$, $V$, $R$, $I$, $z$, $W1$, $W2$, and $W3$ wavebands (see Table\,\ref{tab:RZCep}). However, due to low galactic latitude ($b = 5.5^\circ$) and high $E(B-V)_{\rm SF} = 0.9054 \pm 0.0148$\,mag, this star was excluded from the period--magnitude relation fits presented in Section\,\ref{derivations}. 

\begin{table}
 \centering
% \begin{minipage}{3 in}
  \caption{Mean-flux magnitude data for RZCep.}
  \begin{tabular}{c r}
  \hline

$U$ & $10.4516 \pm 0.0052$ \\
$B$ & $9.7888 \pm 0.0047$ \\
$hipp$ & $9.4727 \pm 0.0075$ \\
$V$ & $9.2304 \pm 0.0123$ \\
$R$ & $9.0369 \pm 0.0090$ \\
$I$ & $8.8288 \pm 0.0166$ \\
$z$ & $9.1478 \pm 0.0260$ \\
$W1$ & $7.8482 \pm 0.0035$ \\
$W2$ & $7.8634 \pm 0.0028$ \\
$W3$ & $7.7455 \pm 0.0047$ \\
\hline
\label{tab:RZCep}
\end{tabular}
% \end{minipage}
\end{table}

Estimating the distance to RZCep using the period--magnitude relations is an excellent test of the results because an HST parallax measurement, ${\mu_{\it HST}=8.02\pm0.17}$, was published as part of \cite{2011AJ....142..187B}.

To apply the period--magnitude relations and fit for a distance modulus to RZCep, Equation\,\ref{eqn:general_PLR} can be rearranged to place the new likelihood information (now including the period--magnitude relation zero point and slope terms) on the left hand side and the formula can be simplified to apply only to RZCep ($i$ subscripts are dropped). The form of the model used for estimating the distance to a single star is thus
\begin{equation}
\label{eqn:RZCep_dist_eqn}
m_{j}  - M_{0,j} - \alpha_j \log_{10}\left(P/P_0\right) =  \mu + E(B-V)\left( a_j R_V + b_j \right) + \epsilon_j,
\end{equation}
where now $\epsilon_j$ is a zero-mean Gaussian random deviate with variance
$$\sigma_{m_j}^2 + \sigma_{M_{0,j}}^2 + \left[\sigma_{\alpha_{j}} \log_{10}\left(P/P_0\right)\right]^2 + \sigma_{j, {\rm intrinsic}}^2.$$

A Bayesian linear regression is fit to solve for the two unknowns, $\mu$ and  ${E(B-V)}$, and $R_V=3.1$ is again used as the extinction law factor (c.f. Section\,\ref{derivations}). The prior distributions should be uninformative and wide [for example, $\mu \sim {\mathcal U}(0, 14)$ and $E(B-V) \sim {\mathcal U}(0, 2)$]. The fit can proceed with mean-flux magnitude measurements in only two bands, but obviously additional waveband data will improve the distance prediction accuracy. 

For RZCep, applying the period--magnitude relations derived in Section\,\ref{derivations} with this Bayesian prediction procedure results in a distance modulus estimate of ${\mu_{\rm PLR} = 8.0397 \pm 0.0123}$ (or ${405.4 \pm 2.3}$ pc). This is a fractional prediction distance error of 0.57 per cent, an improvement of $\sim$13 times the reported HST parallax distance precision \citep{2011AJ....142..187B} and nearly equal to the 14 microarcsec parallax precision (0.57 per cent fractional distance error) Gaia is expected to achieve for bright stellar sources in its end-of-mission analysis \citep{2012Ap&amp;SS.341...31D}.

\begin{figure}
	\centering
	\includegraphics{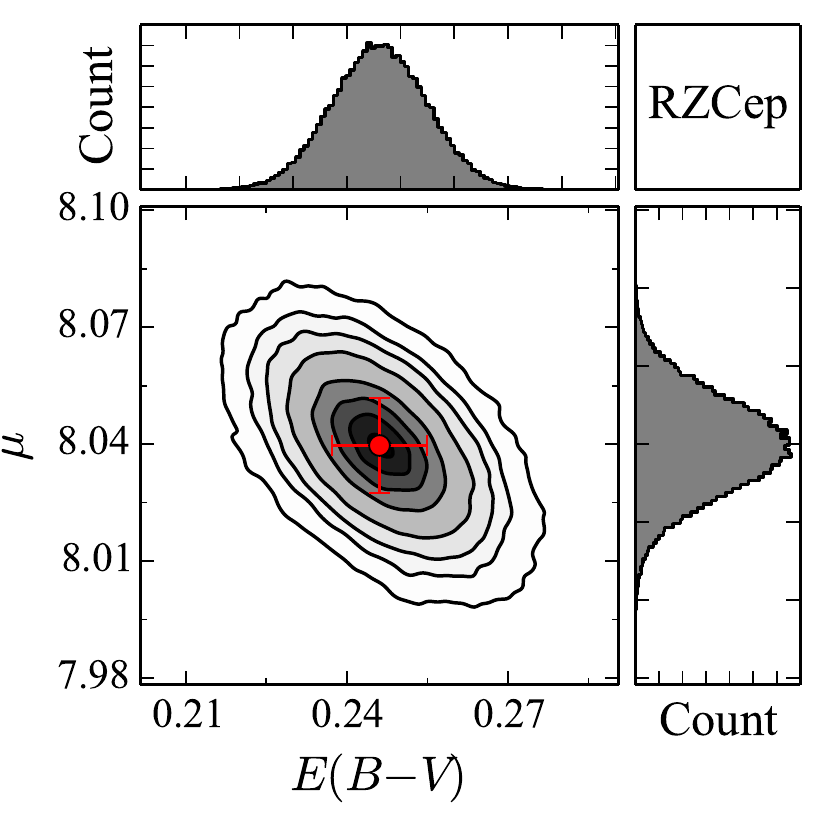}
	\caption{Contour density plot and histograms for the predicted colour excess,  ${E(B-V)}$, and distance modulus, $\mu$, of RZCep from the multi-band period--magnitude relation. 100,000 samples were generated after the MCMC chain converged. The red circle with associated error bars shows the means and standard deviations of the  ${E(B-V)}$ and $\mu$ distributions. Here we fix $R_V = 3.1$.}
	\label{fig:RZCep_contour}
\end{figure}

In solving for the distance prediction, the fit also produces a posterior colour excess value for RZCep, ${E(B-V)_{\rm RZCep} = 0.2461 \pm 0.0089}$. This is significantly less than the line of sight to infinite distance colour excess of ${E(B-V)_{\rm SF} = 0.9054}$, and is very much consistent with RZCep lying only about 400 pc away, even if it is only $5.5^\circ$ off the Galactic plane. This example demonstrates that the multi-band period--magnitude relation can be used to accurately simultaneously fit for an RR Lyrae star's colour excess and distance modulus using only its period and mean-flux magnitude measurements. Fig.\,\ref{fig:RZCep_contour} is the contour density plot for the predicted colour excess and distance modulus for RZCep. The anti-correlation is as expected; for a given brightness, a larger colour excess value requires that the star be closer, and vice versa.

\begin{figure}
	\centering
	\includegraphics{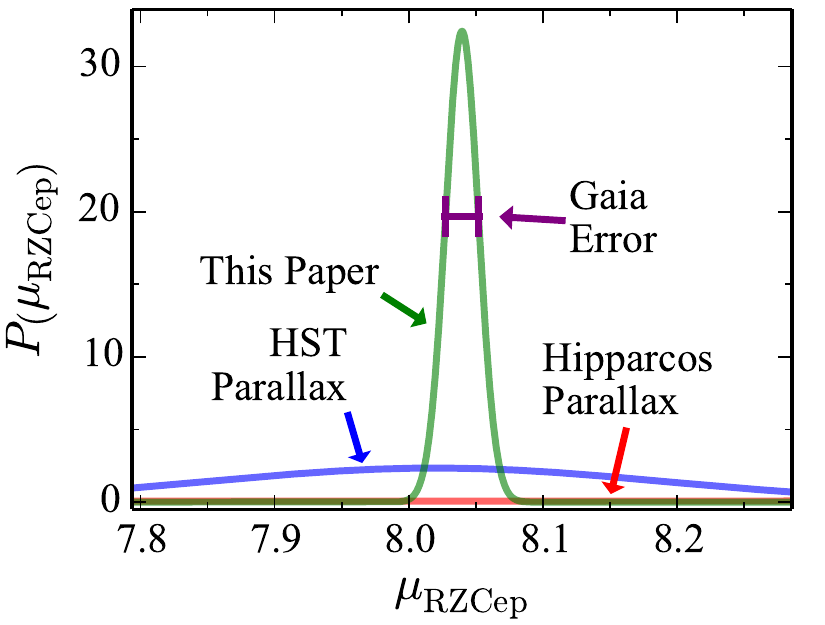}
	\caption{Probability density for the distance modulus of RZCep as determined by various methods. The {\it Hubble Space Telescope} (HST) parallax measurement is from \protect\cite{2011AJ....142..187B}, the {\it Hipparcos} parallax measurement is from \protect\cite{2007ASSL..350.....V}, and the Gaia distance prediction error is 14 microarcsec \citep{2012Ap&amp;SS.341...31D}, which is very similar to the uncertainty on $\mu_{\rm PLR}$ derived in Section\,\ref{application}.}
	\label{fig:RZCep_mu_prediction}
\end{figure} 

\begin{figure}
	\centering
	\includegraphics{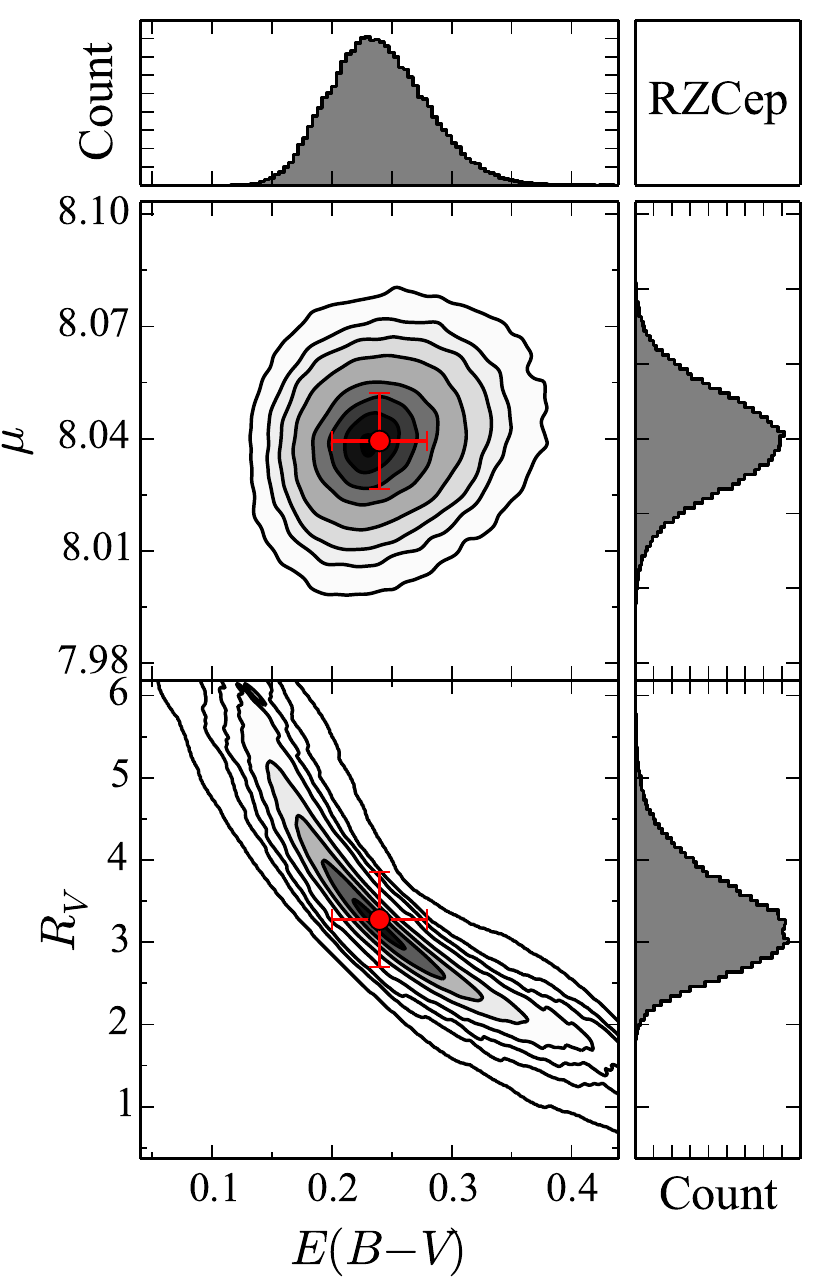}
	\caption{Contour density plot and histograms for the predicted colour excess,  ${E(B-V)}$, extinction law factor, $R_V$, and distance modulus, $\mu$, of RZCep from the multi-band period--magnitude relation with $R_V$ fitted as a model parameter (instead of adopting ${R_V=3.1}$, as was done for the model that produced Fig.\,\ref{fig:RZCep_contour}). 100,000 samples were generated after the MCMC chain converged. The red circle with associated error bars shows the means and standard deviations of the  ${E(B-V)}$, $R_V$, and $\mu$ distributions.}
	\label{fig:RZCep_RV_contour}
\end{figure}

In theory, the fitted model of Equation\,\ref{eqn:RZCep_dist_eqn} can be modified to also fit for the extinction law factor, $R_V$. Such a model was constructed and fit, with the prior distribution of $R_V = {\mathcal N}(3.1, 1)$. The posterior distance is essentially unchanged: ${\mu_{\rm PLR} = 8.0394 \pm 0.0128}$ (or ${405.4 \pm 2.4}$ pc). The posterior colour excess is similar, but substantially wider: ${E(B-V)_{\rm RZCep} = 0.2398 \pm 0.0399}$. And, the posterior $R_V$ is highly covariant with colour excess and very wide: ${R_{V{\rm ,RZCep}} = 3.2768 \pm 0.5820}$. Fig.\,\ref{fig:RZCep_RV_contour} shows the contour density plots for these posterior distributions. Unlike in Fig.\,\ref{fig:RZCep_contour}, the colour excess and distance modulus are not apparently anti-correlated, suggesting that it is effectively the overall magnitude of the bandpass-dependent extinction (set by the combination of ${E[B-V]}$ and $R_V$) which is most directly constrained by the data.

\section[]{Discussion and Conclusions}\label{conclusions}

We have applied a simultaneous Bayesian linear regression methodology to 637 mean-flux magnitude measurements of a calibration sample of 134 RR Lyrae stars to derive new, tightly-constrained RR Lyrae period--magnitude relations in 13 wavebands. As part of the regression model, the colour excess, ${E(B-V)}$, for each star was also determined. The final result is that the distances to the 134 calibration stars are measured with median fractional error of 0.66 per cent. We showed how the period--magnitude relations can be used singly or in combination through the methodology described in Section\,\ref{application} to derive distances to other observed RR Lyrae stars achieving a similar level of precision.

As part of the multi-band fit, the intrinsic scatter, $\sigma_{\rm intrinsic}$, for each period--magnitude relation was constrained. Intrinsic scatter is the residual about the best-fit period--magnitude relation in each waveband which cannot be accounted for by instrumental photometric error. It was found that $\sigma_{\rm intrinsic}$ is minimised for the mid-infrared $W1$ and $W2$ wavebands, indicating that this wavelength regime provides the most accurate absolute magnitude predictions via its period--magnitude relations. The overall principle of the simultaneous multi-band fitting methodology is that additional wavelength data for a star is always useful in improving the absolute magnitude (and thus distance) estimate accuracy. However, the $\sigma_{\rm intrinsic}$ analysis suggests that the most valuable wavebands for this purpose are around 3-5\,$\mu$m.

Also of note are the specific results pertaining to fitted colour excess for each calibration RR Lyrae star. The regression found that tighter period--magnitude relation fits were possible by preferring a systematic increase in posterior colour excess, as compared to the $E(B-V)_{\rm SF}$ values (c.f. Fig.\,\ref{fig:ebv_residual}). While this could be explained by a systematic error in the \cite{1998ApJ...500..525S} dust map or an $R_V$ value significantly different from 3.1, a possible physical explanation is that RR Lyrae stars are often enshrouded in a local dust envelope. Stellar evolution models require the progenitors of RR Lyrae stars to shed about ${0.1~M_\odot}$ of material within a few hundred million years as the stars evolve to the horizontal branch \citep{1995CAS....27.....S}. This material, blown off the stars via stellar wind, may cool to form dust. Strict conclusions should not be drawn from the findings of these period--magnitude relation fits, but further investigation of this hypothesis is encouraged.

As an alternative method for presenting the results, the period--magnitude relations were used to calculate the mean-flux SEDs of RR Lyrae stars as a function of period, as presented in Fig.\,\ref{fig:seds}. The derived SEDs coincide with model stellar spectra calculated at bounding values of RR Lyrae effective temperature and radius. This perspective view of the period--magnitude relations makes clear why the relation slope, $\alpha$, is observed to asymptote at longer wavelengths, as shown in Fig.\,\ref{fig:slope}.

There are some possible extensions to the formalism we have presented. First, in the analysis presented we fixed $R_V = 3.1$ but $R_V$ could be left as a free parameter either globally or for every line of sight. In our initial work for this paper we left $R_V$ to be free for every line of sight and found that the MCMC chains did not converge. This is understandable given the degeneracies between $\mu$, $R_V$ and ${E(B-V)}$ in the formulation and the fact that many stars in the sample had only a few bandpasses in which mean magnitudes were measured. Nevertheless, we believe that $\sigma_{\rm intrinsic}$ serves to capture any of the potential systematic errors that might be induced by a variable or globally different value of $R_V$. There is some validation on this point in that when we allowed $R_V$ to vary for RZCep, traces in $R_V$ and  ${E(B-V)}$ are strongly anti-correlated but the inferred distance posterior is essentially unchanged (see Fig.\,\ref{fig:RZCep_RV_contour}). By adding UV data (say from the Swift satellite), in a future work, the degeneracy between  $R_V$ and ${E(B-V)}$ may be broken.  Second, we did not include in the formalism any term related to a possible effect on metallcity, lacking a physically motivated parametrization for doing so. If such a theoretical formalism is found, it could be easily incorporated. Note that we found no correlation of deviations from period-magnitude relations with metallicity, again offering $\sigma_{\rm intrinsic}$ as the likely capturer of any systematic errors of unmodelled metallicity dependencies.

The future applications of the derived RR Lyrae period--magnitude relations range from nearby Milky Way structure studies to distance measurements at truly cosmic scales (pushing into the Hubble Flow at $d>100$ Mpc). Ground-based optical surveys (PanSTARRS, iPTF, Catalina Sky Survey, OGLE IV, LSST, etc.) and the proliferation of near-infrared followup facilities (RATIR, NEWFIRM, UKIRT, etc.) are now enabling studies of Milky Way Field and Halo RR Lyrae stars to produce highly accurate distance measurements. Mid-infrared facilities and surveys (SOFIA, {\it Spitzer Space Telescope}, MaxWISE, and in the near future, JWST) can also be leveraged to significantly improve RR Lyrae distance measurement precision. These studies will use the RR Lyrae period--magnitude relations to map Milky Way stellar density, measure the morphology of remnant tidal streams in the Halo, and probe the depth structure of the Magellanic Clouds. 

Additionally, as demonstrated in the present work, combining optical and infrared light curve data for an RR Lyrae star can provide a fit for both distance and colour excess along that line of sight to that distance. Given enough RR Lyrae targets (\citealt{2012Ap&amp;SS.tmp...49E} predicts $\sim$100,000 RR Lyrae stars in the Milky Way), a 3D dust map can be constructed to better understand the distribution of Milky Way dust grains and to also aide in estimating line-of-sight extinction for studies of other objects within the Milky Way.  As a cross check and calibrator, we see precision 3D line-of-sight dust measurements \citep{2011MNRAS.411..435B} as complementary to the ongoing all-sky efforts using aggregate stellar populations \citep{2012MNRAS.427.2119S,2012ApJ...757..166B,2014MNRAS.438.2938H,2014ApJ...783..114G}, which offer aggregate dust measures over arcminute scales and in wide distance bins. With a significantly larger sample, it will be also possible to test how universal the power-law fits are for different subpopulations of RR Lyrae: there may very well be measurable differences in relations as a function of metallicity, environment and/or population origin (e.g., thick disk vs.\ bulge). 

RR Lyrae stars serve as primary distance indicators in the Cosmic Distance Ladder via their period--magnitude relations. As such, RR Lyrae stars are vital to calibrating the relations used for secondary distance indicators that extend out well beyond the Local Group. Error in distance measurement methods propagates up the distance ladder, and thus minimisation of error at the local end can significantly improve the accuracy of secondary indicators and the derived higher-level measurements, such as $H_0$. This effect, as applied through improving the Cepheid Leavitt Law to better constrain Type Ia supernovae luminosity, has recently been very well utilised by both \cite{2011ApJ...730..119R} and \cite{2012ApJ...758...24F} in their measurements of $H_0$ with $\sim3$ per cent precision.

RR Lyrae stars, in combination with the TRGB method to reach distant supernova host galaxies, offer a systematically separate and competitive means for Type Ia supernova luminosity calibration. Additional physical distance measurement methods such as this are necessary to help resolve the conflict between the $H_0$ values found by the distance ladder methods of \cite{2011ApJ...730..119R} (${73.8 \pm 2.4 ~ {\rm km}~{\rm s}^{-1}~{\rm Mpc}^{-1}}$) and \cite{2012ApJ...758...24F} (${74.3 \pm 2.1 ~ {\rm km}~{\rm s}^{-1}~{\rm Mpc}^{-1}}$), and the statistically significantly lower measurement derived by \cite{2013arXiv1303.5076P} with Cosmic Microwave Background data from the Planck satellite (${67.3 \pm 1.2 ~ {\rm km}~{\rm s}^{-1}~{\rm Mpc}^{-1}}$). If our methodology holds up to scrutiny the achievement of sub-1 per cent fractional distance errors (herein, 0.66 per cent for the calibration sample) within a formalism that accounts for dust extinction may be considered a strong buttressing of the path to eventual 1 per cent uncertainties in Hubble's constant.

\section*{Acknowledgments}

The authors acknowledge the generous support of grants (\#0941742 and \#1009991) from the National Science Foundation. The authors acknowledge and thank Ed Rosten for providing the normalised cross correlation image-alignment program utilised as part of the {\it PAIRITEL} reduction pipeline. {\it PAIRITEL} was operated by the Smithsonian Astrophysical Observatory (SAO) and was made possible by a grant from the Harvard University Milton Fund, a camera loan from the University of Virginia, and continued support of the SAO and UC Berkeley.  The research with {\it PAIRITEL} was also support by NASA guest investigator grants {\sc NNX12AE67G} and {\sc NNX13AC58G}. We are grateful for the assistance of the staffs at all of the observatories used to obtain the data. This research has made use of the NASA/IPAC Infrared Science Archive, which is operated by the Jet Propulsion Laboratory, California Institute of Technology, under contract with the National Aeronautics and Space Administration.
This publication makes use of data products from the Two Micron All Sky Survey, which is a joint project of the University of Massachusetts and the Infrared Processing and Analysis Center/California Institute of Technology, funded by the National Aeronautics and Space Administration and the National Science Foundation.
This publication makes use of data products from the Wide-field Infrared Survey Explorer, which is a joint project of the University of California, Los Angeles, and the Jet Propulsion Laboratory/California Institute of Technology, funded by the National Aeronautics and Space Administration. This publication also makes use of data products from NEOWISE, which is a project of the Jet Propulsion Laboratory/California Institute of Technology, funded by the Planetary Science Division of the National Aeronautics and Space Administration. This research has made use of NASA's Astrophysics Data System. 

\bibliographystyle{mn2e_alt}
\bibliography{Klein_refs}

%\bsp

\label{lastpage}

\end{document}